\pdfoutput=1
\documentclass[prd,twocolumn,superscriptaddress,preprintnumbers,nofootinbib]{revtex4-2}
\usepackage{graphicx}
\usepackage{epsfig}
\usepackage{bm}
\usepackage{latexsym,amssymb,amsmath,amsfonts,amssymb,txfonts,wasysym,float}
\usepackage{scrextend}
\usepackage{mathrsfs}
\usepackage{mathtools}
\usepackage{color}
\usepackage[caption=false]{subfig}

\usepackage{enumitem}
\usepackage{orcidlink}
\definecolor{colorLink}{rgb}{0,0,180}
\usepackage{hyperref}
\hypersetup{
   colorlinks = true,
   citecolor  = colorLink,
   urlcolor   = colorLink,
   linkcolor  = colorLink,
}

\newcommand{\xmax}{X_{\text{max}}}

\DeclareMathOperator{\sgn}{sgn}
\DeclarePairedDelimiter\abs{\lvert}{\rvert}%

\usepackage{xcolor}
\definecolor{orange}{cmyk}{0,0.5,1,0}
\definecolor{rossoCP3}{cmyk}{0,.88,.77,.40}
\definecolor{graa}{rgb}{0.8,0.8,0.8}
\definecolor{blaa}{rgb}{0.2,0.2,0.6}

\usepackage[sort&compress]{natbib}

\begin{document}

\title{\color{rossoCP3}  A Peters cycle at the end of the cosmic ray spectrum?}

\author{\bf Marco Stein Muzio \orcidlink{0000-0003-4615-5529}}

\email{msm6428@psu.edu}

\affiliation{Department of Astronomy and Astrophysics, Pennsylvania
  State University, University Park, PA 16802, USA}
\affiliation{Department of Physics, Pennsylvania State University, University Park, PA 16802, USA}

\affiliation{Institute of Gravitation and the Cosmos, Center for Multi-Messenger Astrophysics, Pennsylvania State University, University Park, PA
16802, USA}

\author{\bf Luis A. Anchordoqui
\orcidlink{0000-0003-1463-7136}}

\affiliation{Department of Physics and Astronomy,  Lehman College, City University of
  New York, NY 10468, USA
}

\affiliation{Department of Physics,
 Graduate Center, City University
  of New York,  NY 10016, USA
}

\affiliation{Department of Astrophysics,
 American Museum of Natural History, NY
 10024, USA
}

\author{\bf Michael Unger \orcidlink{0000-0002-7651-0272}}
\affiliation{Institut f\"ur Astroteilchenphysik, Karlsruher Institut f\"ur Technologie, Karlsruhe 76344, Germany}
\affiliation{Institutt for fysikk, Norwegian University of Science and Technology (NTNU), Trondheim, Norway}

\date{\today}

\begin{abstract}

\noindent We investigate the degree to which current ultrahigh energy cosmic ray observations above the ankle support a common maximum rigidity for all nuclei, often called a Peters cycle, over alternative scenarios for the cosmic ray spectra escaping sources. We show that a Peters cycle is not generally supported by the data when compared with these alternatives for the scenarios studied. We explore the observational signatures of non-Peters cycle scenarios, and the opportunities to explore both ultrahigh energy cosmic ray source conditions, as well as, physics beyond the Standard model they present.

\end{abstract}
\maketitle

\section{Introduction}\label{sec:1}
One of the most challenging open questions regarding the origin of ultrahigh energy cosmic rays (UHECRs) deals with the relative maximum energies of the spectra escaping the source for different nuclei.
A commonly-used simplifying assumption is that the cosmic ray source spectra trace a Peters cycle, in which the maximum cosmic ray energy scales linearly with $Z$, i.e., with the charge of the UHECR in units of the proton charge~\cite{Peters:1961} (see e.g.~\cite{Allard:2005cx,Taylor:2013gga,PierreAuger:2016use} for fits of UHECR data with a Peters cycle at the source). A Peters cycle arises naturally for acceleration processes which depend on magnetic fields to confine cosmic rays to the acceleration region. This limits the maximum rigidity of the accelerator to $R_\mathrm{max} \lesssim BL$, where $B$ and $L$ are the magnetic field strength and size of the accelerating region. This condition, often called the Hillas criterion~\cite{Hillas:1984ijl}, results in a common maximum rigidity among all nuclei which are accelerated.

Current UHECR data, though, is not compatible with a pure Peters cycle from a single source population across the entire spectrum~\cite{Luce:2022awd,PierreAuger:2022atd} above $10^{18}$~eV. In particular, current spectrum and composition data require at least one of two possibilities in order to be reconciled. The first possibility is that the UHECR flux below the ankle is dominantly produced by a second class of extragalactic sources (see~\cite{PierreAuger:2022atd}). However, the number of different source populations and their relative variance cannot be too large~\cite{Ehlert:2022jmy}.
Another possibility is that an alternative scaling of maximum energies is required to explain the full UHECR spectrum and composition data (see, e.g.,~\cite{Unger:2015laa,Globus:2015xga,Kachelriess:2017tvs,Muzio:2019leu}). Such alternative scalings are a natural assumption for models which consider the possibility that UHECRs escape their source only after suffering significant energy losses. A specific illustration of this in the context of gamma-ray bursts (GRBs) is presented in~\cite{Biehl:2017zlw}.

However, since the CR escape time naturally decreases with rigidity, it is possible that the highest energy CRs suffer minimal energy losses and escape their sources with a Peters cycle intact. Therefore, here we consider whether the UHECR data above the ankle, in particular above $10^{18.8}$~eV, can still be reconciled with a pure Peters cycle, rather than one modified by energy losses.

Energy losses are typically parametrized in terms of $Z$ and the UHECR baryon number $A$. For example, the energy loss rate of synchrotron and curvature radiative processes scales as
$Z^4/A^{2}$ and $Z^2$, respectively, see e.g.~\cite{Anchordoqui:2018qom}, leading to different scalings for the maximum energy~\cite{Medvedev:2003sx,Ptitsyna:2008zs,Rieger:2011ch}. For instance, when considering a diffusive acceleration mechanism, synchrotron radiation limits the cosmic ray maximum energy so that it scales as $E_{\rm max} \propto A^4/Z^4$.
On the other hand, for one-shot acceleration processes, synchrotron radiation constrains the maximum energy to scale as $E_{\rm max} \propto A^2/Z^{3/2}$, whereas curvature radiation yields $E_{\rm max} \propto A/Z^{1/4}$. Photonuclear
interactions with the thermal radiation fields and hadronic interactions
with the ambient gas are driven by the scattering cross section, which scales with $A$. Even consideration of a single
particle species at injection can provide a complex multi-species source spectra
after propagation through the source environment~\cite{Unger:2015laa,Muzio:2019leu,Muzio:2021zud}.

The maximum energy of cosmic rays may even be independent of both $Z$ and $A$. A specific example of this is a beyond the Standard Model (BSM) dark dimension scenario, which naturally addresses the cosmological hierarchy problem by adding one mesoscopic dimension of micron scale~\cite{Montero:2022prj}. Since within this scenario physics becomes strongly coupled to gravity around $10^{10}~{\rm GeV}$, it was recently conjectured that new universal energy losses deep-rooted within the dark dimension could control the cosmic ray maximum energy~\cite{Montero:2022prj,Anchordoqui:2022ejw,Noble:2023mfw}.

In this paper we take a pragmatic approach to investigate whether existing data favor any of the above mentioned scenarios over a pure Peters cycle above the ankle. Using data from the Pierre Auger Observatory, we carry out a statistical analysis to study the degree to which the observed spectrum and nuclear composition constrain the source emission spectra.

The layout of the paper is as follows. In Sec.~\ref{sec:2} we provide an overview of the data and we introduce the framework for a data-driven analysis to model the maximum cosmic ray energy and constrain the source spectra. In Sec.~\ref{sec:3} we discuss the results for benchmark models. We find that a Peters cycle is not the most favorable model. In Sec.~\ref{sec:4} we provide an astrophysical interpretation of our results and we discuss possible  multimessenger connections and implications.  Our conclusions are collected in Sec.~\ref{sec:5}.

\section{Model}
\label{sec:2}
We adopt the working assumption that the all source spectra can be described by
\begin{align}
    \frac{dN(A)}{dE} \varpropto E^{-\gamma} \, e^{-E/E^A_{{\rm max}}} \,,
\end{align}
with
\begin{equation}
 E_{{\rm max}}^A = E_0 \, Z^\alpha A^\beta,
\end{equation}
where $E_0$ is the proton maximum energy of the system, and the spectral index is common to all nuclei. Note that the case with $\alpha =1$ and $\beta = 0$ corresponds to the Peters cycle, whereas the case with $\alpha =0$ and $\beta = 0$ stands for the case in which the spectrum is dominated by some universal energy loss which is independent of the cosmic ray Lorentz boost.

\par
With this model for the spectra escaping the source we consider five mass groups representing $p$, He, CNO, Si, and Fe, whose relative abundances are adjusted to obtain the best-fit. The mass groups are propagated through the cosmic microwave background (CMB) and extragalactic background light (EBL)~\cite{Gilmore:2011ks} using propagation matrices built from CRPropa3~\cite{AlvesBatista:2016vpy} (which assumes \textsc{Talys}~v1.8~\cite{Koning:2005ezu} for photodisintegration cross-sections). We consider sources to a maximum comoving distance of $4.2$~Gpc and assume they follow a star-formation rate (SFR) evolution~\cite{Robertson:2015uda}. Additionally, we use The observed spectrum of CRs is fit to the spectrum~\cite{PierreAuger:2021hun} and composition~\cite{Yushkov:2020nhr} of the Pierre Auger Observatory (Auger) by minimizing the value of a combined $\chi^2$,
\begin{align}
    \chi^2 = \displaystyle \sum_i \frac{(J_i-\hat{J}_{i})^2}{\sigma_{J,i}^2} + \sum_i \frac{(\mu_i-\hat{\mu}_{i})^2}{\sigma_{\mu,i}^2} + \sum_i \frac{(V_i-\hat{V}_{i})^2}{\sigma_{V,i}^2} ~,
\end{align}
where $J$ is the UHECR flux and $\mu$ and $V$ are the mean and
variance of the distribution of the logarithm of cosmic ray mass
number, $\ln{A}$. Quantities with a hat are the model prediction and
the subscript $i$ denotes the energy. In particular, to interpret the
air shower data provided by Auger we must adopt a hadronic interaction
model. We consider two hadronic interaction
models, \textsc{Sibyll2.3d}~\cite{Riehn:2019jet}
and \textsc{Epos-LHC}~\cite{Pierog:2013ria}, to interpret the depth of
shower maximum, $X_\mathrm{max}$, data in terms of $\ln{A}$.
It is worthwhile noting that in general the first two moments of the $X_\mathrm{max}$ distribution do not contain its full information (unless the distribution is Gaussian). We accept this loss of information in the fit due to (a) the simplicity of fitting just the moments and (b) the larger data set for which data on the moments is available in comparison to the full distribution ($20\%$ more statistics and one additional energy bin above $10^{18.8}$~eV).

\par
We perform a fit to the Auger data~\cite{PierreAuger:2021hun,Yushkov:2020nhr} above $10^{18.8}$~eV so as to directly address the question of whether a pure Peters cycle is preferred above the ankle. At lower energies, we assume either an additional source population contributes to the spectrum or that energy loss processes have significantly distorted the original Peters cycle. Given our free parameters (the proton maximum energy $E_0$, spectral index $\gamma$, and $4$ parameters controlling $5$ mass group fractions) this choice of fit range leaves $N_\mathrm{dof}=29$. Here we focus on a benchmark set of systematic data shifts, which provide the best-fit to the Auger data overall: shifting the energy scale by $\mathrm{dlg}E=+0.1$ and shifting the $\langle X_\mathrm{max} \rangle$ by $-1\sigma_X$, since these gave the best overall fit of the shifts explored. We consider the sensitivity of our results to other systematic shifts of the data in Appendix~\ref{app:sysShifts}.

\section{Results}
\label{sec:3}

\par
After optimizing the model parameters for a particular $(\alpha,\beta)$ combination we calculate the relative goodness-of-fit, $N_\sigma$, compared to a Peters cycle in units of sigma. To calculate $N_\sigma$ we apply Wilks' theorem to convert the $\Delta \chi^2$ between an alternative scenario and a Peters cycle to a p-value, given that alternative scenarios have $\Delta N_\mathrm{dof} = 2$. We calculate $\Delta \chi^2$ as
\begin{align}
     \Delta \chi^2 \equiv S^{-1} \sqrt{\abs*{\chi_{\alpha,\beta}^2 - \chi^2_\mathrm{Peters}}}
\end{align}
where $S = \smash[b]{\left(\min(\chi_{\alpha,\beta}^2,\chi^2_\mathrm{Peters})/N_\mathrm{dof})\right)}^{1/2}$ is a scale factor introduced to enlarge the uncertainties to account for a $\chi^2_\mathrm{min}/N_\mathrm{dof} > 1$~\cite{Rosenfeld:1975fy}. Additionally, we assign $N_\sigma$ a sign equal to $\sgn\left(\chi_{\alpha,\beta}^2 - \chi^2_\mathrm{Peters}\right)$ to encode whether the fit has improved or worsened compared to a Peters cycle. In other words, negative values of $N_\sigma$ represent the statistical significance at which one can reject the null hypothesis of a pure Peters cycle in favor of the alternative scenario $(\alpha, \beta)$.

\begin{figure*}[htbp!]
    \centering
    \begin{minipage}{0.49\linewidth}
        \centering
        \subfloat[\label{fig:sibyll_benchmark}]{\includegraphics[width=\textwidth]{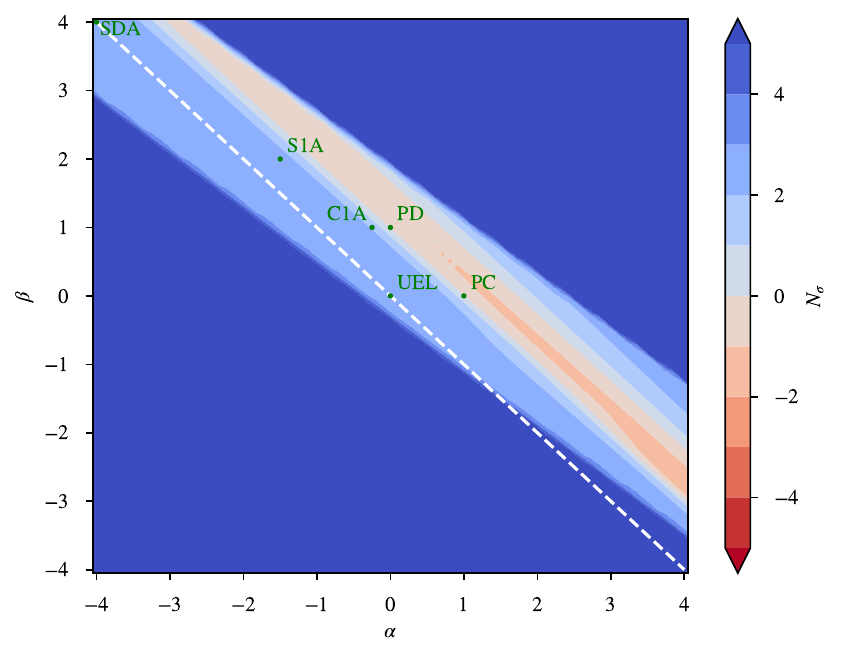}}
    \end{minipage}
    \begin{minipage}{0.49\linewidth}
	  \centering
        \subfloat[\label{fig:epos_benchmark}]{\includegraphics[width=\textwidth]{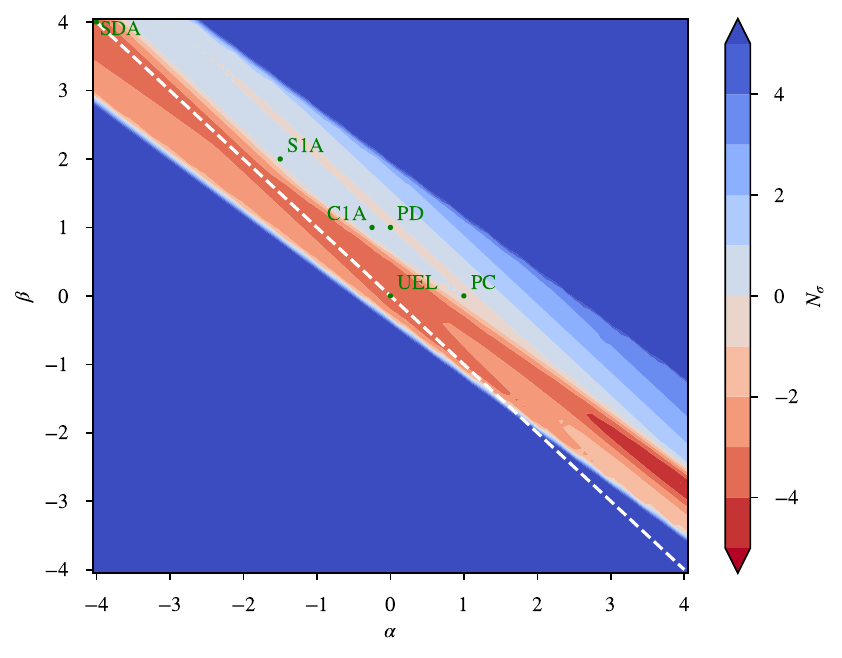}}
    \end{minipage}
	\caption{Change in quality of fit to the UHECR spectrum and composition relative to a Peters cycle. We consider (a) \textsc{Sibyll2.3d} and (b) \textsc{Epos-LHC} with data shifted by $\mathrm{dlg}E=0.1$ and $-1\sigma_X$. The family of scenarios with $\alpha + \beta = 0$ are indicated by the white dashed line. The Peters cycle (PC) and a number of alternative scenarios (green dots) are highlighted: a photodisintegration-limited spectrum (PD), a synchrotron-limited diffusion accelerated spectrum (SDA), a synchrotron-limited one-shot accelerated spectrum (S1A), a curvature radiation-limited one-shot accelerated spectrum (C1A), and a universal energy loss spectrum (UEL).}
	\label{fig:benchmark_models}
\end{figure*}

\par
Figure~\ref{fig:benchmark_models} shows $N_\sigma$ for \textsc{Sibyll2.3d} and \textsc{Epos-LHC} using our benchmark systematic shifts for the data. For reference, the Peters cycle (PC, $\alpha=1$, $\beta=0$) and a number of alternative scenarios discussed in the Section~\ref{sec:1} are highlighted: a photodisintegration-limited spectrum (PD, $\alpha=0$, $\beta=1$), a synchrotron-limited diffusion accelerated spectrum (SDA, $\alpha=-4$, $\beta=4$), a synchrotron-limited one-shot accelerated spectrum (S1A, $\alpha=-3/2$, $\beta=2$), a curvature radiation-limited one-shot accelerated spectrum (C1A, $\alpha=-1/4$, $\beta=1$), and a universal energy loss spectrum (UEL, $\alpha=0$, $\beta=0$) as would be predicted for a BSM dark dimension scenario.

\par
The value of $\alpha$ and $\beta$ change the relative position of nuclei in the spectrum emerging from the source (as illustrated in Appendix~\ref{app:esc_spectra}). This in turn changes the relative position of mass groups at Earth which affects the quality of fit to the CR composition data. Therefore, one expects that points in the $\alpha-\beta$ plane with similar ratios of maximum energy between nuclei to produce similar quality fits. In practice this is realized due to the approximate degeneracy between $A$ and $Z$, since $A \simeq 2Z$ for stable nuclei with the exception of protons where $A=Z$ (though these fall below our fit range in some cases, including the standard Peters cycle scenario). In particular, the ratio between maximum energies for two nuclei $A$ and $A'$ will be constant along lines of constant $\alpha + \beta$, since:
\begin{align}\label{eq:nucleiRatio}
    \frac{E_\mathrm{max}^A}{E_\mathrm{max}^{A'}} = \left( \frac{A}{A'}\right)^{\alpha+\beta} ~.
\end{align}
By contrast, the ratio between the maximum energies of a nucleus $A$ and protons will be constant along lines of constant $(1-\log_{A}{2})\alpha + \beta$, since:
\begin{align}\label{eq:ApRatio}
    \frac{E_\mathrm{max}^A}{E_\mathrm{max}^{p}} = 2^{-\alpha} A^{\alpha+\beta}~.
\end{align}

Equations~\eqref{eq:nucleiRatio} and~\eqref{eq:ApRatio} have two important consequences. First, in the case that the escaping proton flux is not significant relative to other mass groups, different acceleration scenarios fall into families which have a common $\alpha+\beta$. Within these families, one can expect that the observed UHECR spectrum and composition are nearly indistinguishable despite the fundamentally different processes responsible for them. Second, in the case that a substantial proton population escapes the source, the proton spectrum can peak at arbitrarily high energies compared to the peak energies of nuclei irrespective of the sign of $\alpha+\beta$. As an example, this means that even for scenarios in the Peters cycle family $\alpha+\beta=1$, nuclei remain ordered in terms of their peak energy but protons can peak below helium, above iron, or between the two.\\

As can be seen from Fig.~\ref{fig:benchmark_models}, alternative scenarios are favored over a simple Peters cycle. For \textsc{Epos-LHC} alternative scenarios can achieve $N_\sigma < -4$, indicating that these scenarios could be used to reject the Peters cycle hypothesis with strong statistical significance. In particular, among the highlighted alternative scenarios the UEL case achieves more than $3\sigma$ improvement over a Peters cycle. This directly demonstrates the potential for UHECR data to probe BSM processes if additional statistical significance is added but future data. Notably the region with the most negative value of $N_\sigma$ has a slope deviating from $-1$, indicating that the position of protons relative to heavier nuclei is responsible for the improvement in fit quality. In particular, this region satisfies $\beta \simeq 0.4 - 0.8\alpha$ and we can infer that it is the ratio of the maximum proton energy to the maximum energy of $A\simeq 32$ which drives the fit.

\par
\begin{figure}[htbp!]
    \centering
    \includegraphics[width=\linewidth]{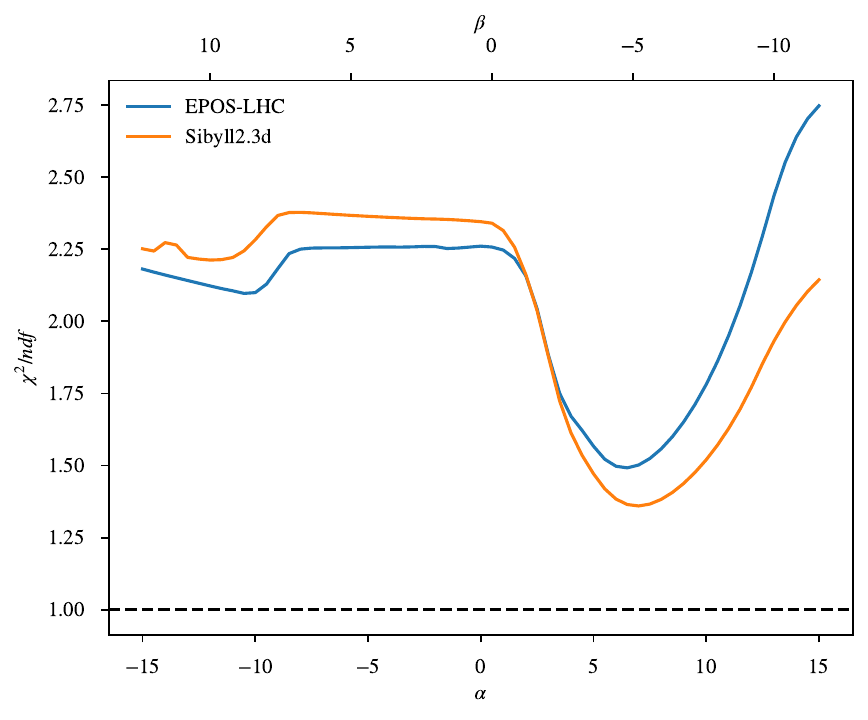}
    \caption{Reduced $\chi^2$ along the line $\beta = 0.4 - 0.8 \alpha$ for \textsc{Sibyll2.3d} and \textsc{Epos-LHC}. In both cases the minimum appears around $\alpha\simeq 6.75$ and $\beta \simeq = -5$.}
    \label{fig:linear_scan}
\end{figure}

It is clear from Fig.~\ref{fig:benchmark_models} that the global minimum is outside the plotted range (which was driven by the range of $\alpha$ and $\beta$ values of the alternative scenarios we considered). To explore how far outside of this range the global minimum lies, we performed a 1-D scan along the line giving the best-fit for \textsc{Epos-LHC}: $\beta \simeq 0.4 - 0.8\alpha$. The results of this scan are shown in Fig.~\ref{fig:linear_scan}. Independent of the hadronic interaction model, the best-fit is found for $\alpha\simeq 6.75$ and $\beta \simeq -5$. Currently, we are unaware of any scenarios which could produce such an exotic dependence on mass and charge.

\par
\begin{figure*}[!t]
    \def\figw{0.68}
    \centering
    \begin{minipage}{\figw\linewidth}
        \centering
        \subfloat[Fit using \textsc{Sibyll2.3d} for the relation between mass and shower maximum, $\xmax$.\label{fig:sibyll_obs}]{\includegraphics[width=\textwidth]{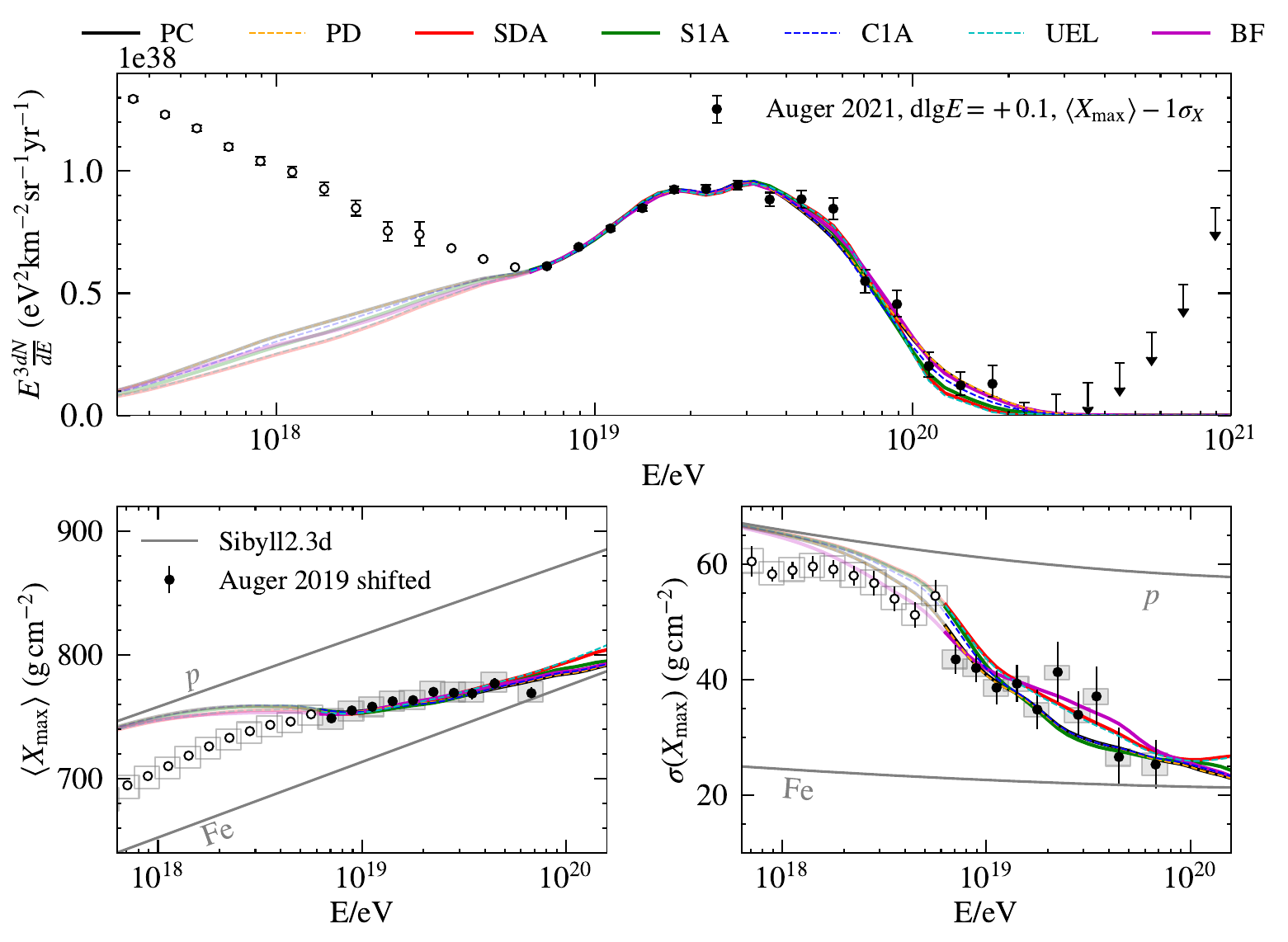}}
    \end{minipage}\\[\baselineskip]
     \begin{minipage}{\figw\linewidth}
        \centering
        \subfloat[Fit using  \textsc{Epos-LHC} for the relation between mass and shower maximum, $\xmax$. \label{fig:epos_obs}]{\includegraphics[width=\textwidth]{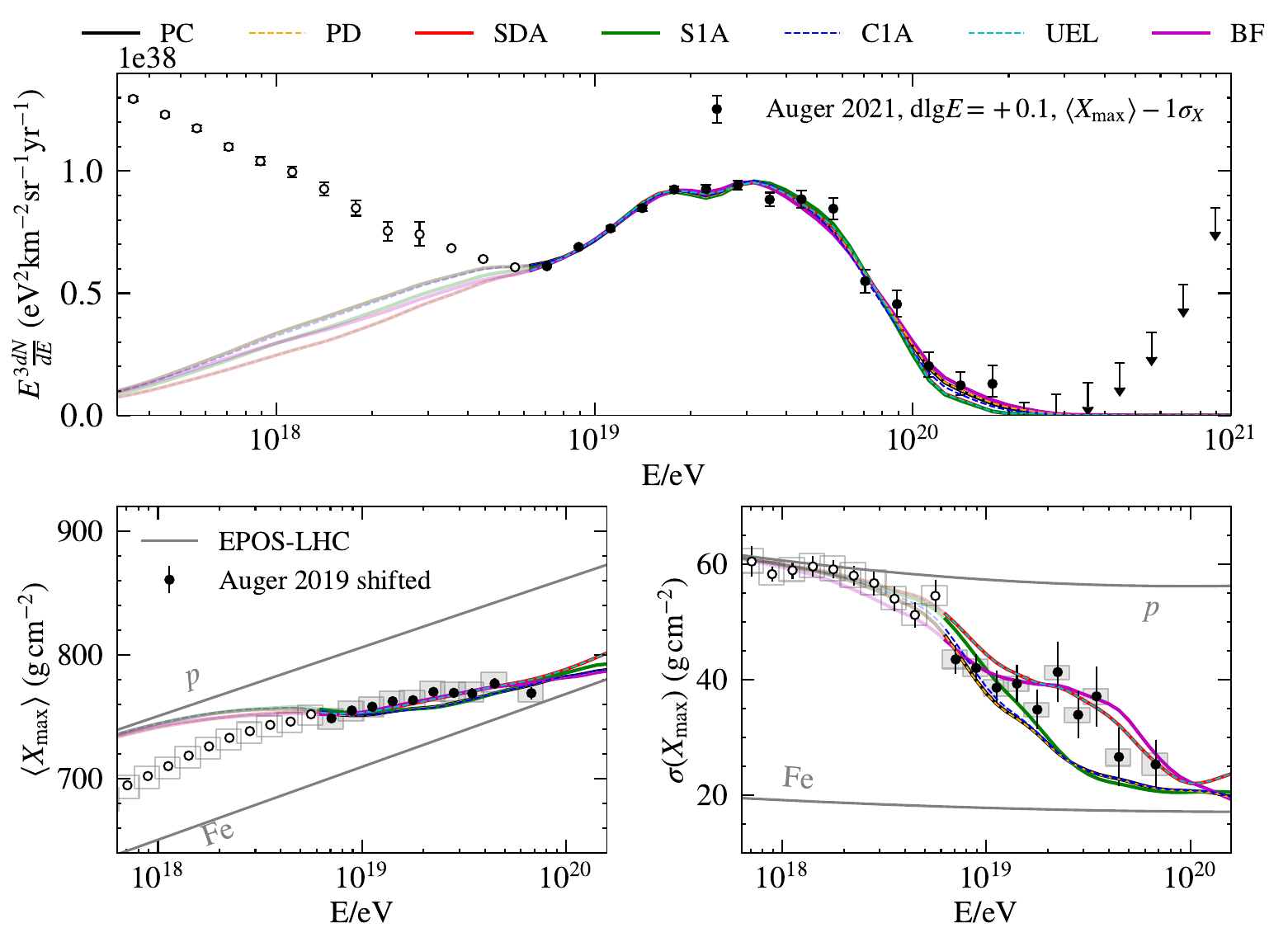}}
    \end{minipage}
    \caption{Best-fit spectra (top panels) and composition-related observables (bottom panels) assuming a Peters cycle and other alternative scenarios (colored lines). Data points are the Auger 2021 spectrum~\cite{PierreAuger:2021hun} and 2019 composition data~\cite{Yushkov:2020nhr} shifted by $\mathrm{dlg}E=+0.1$ and $-1\sigma_X$. Predictions for $\langle X_\mathrm{max} \rangle$ and $\sigma(X_\mathrm{max})$ are made assuming (a) \textsc{Sibyll2.3d} and (b) \textsc{Epos-LHC}. Precitions for pure proton and iron compositions for each hadronic interaction model are shown for reference (solid gray lines). Models were fit to the data above $10^{18.8}$~eV (solid points). Data points below the fit range are shown for reference (open points). The different models are: Peters cycle (PC), a photodisintegration-limited spectrum (PD), a synchrotron-limited diffusion accelerated spectrum (SDA), a synchrotron-limited one-shot accelerated spectrum (S1A), a curvature radiation-limited one-shot accelerated spectrum (C1A), and a universal energy loss spectrum (UEL).
    }
    \label{fig:obs_spectra}
\end{figure*}

Figure~\ref{fig:obs_spectra} shows the best-fit spectrum and composition-related observables under the Peters cycle and alternative scenarios highlighted in Fig.~\ref{fig:benchmark_models}. The global best-fit (BF) scenario found from the 1-D scan discussed above is also plotted in Fig.~\ref{fig:obs_spectra}. While predictions for the UHECR spectrum are very similar between all the scenarios considered, there are differences in their predictions for the UHECR composition.
It is worthwhile noting that the models presented here explain nearly all the flux in the two data points below our fit range, requiring a fine-tuned transition to a second, extragalactic source population. This may hint that the UHECR spectrum is best explained by a energy-loss-modified spectrum, because the nucleons created during energy losses in the source environment can populate the flux just below the ankle, see \cite{Unger:2015laa}.

\par
As is hinted at by Fig.~\ref{fig:benchmark_models} and \eqref{eq:nucleiRatio}, the best-fit spectrum and composition for models with similar $\alpha+\beta$ are nearly indistinguishable from each other if no significant proton component exists in the escaping spectrum. This explains the stark similarity both between a Peters cycle and photodisintegration-limited scenario ($\alpha+\beta=1$), as well as, between a synchrotron-limited diffusion accelerated scenario and a universal energy loss scenario ($\alpha+\beta = 0$). Owing to its having $\alpha+\beta= 0.75$, the curvature radiation-limited one-shot accelerated scenario falls between the Peters cycle/photodisintegration-limited scenario and the synchrotron-limited one-shot accelerated scenario ($\alpha+\beta=0.5$). While some of these alternative scenarios give slightly better or worse fits compared to the Peters cycle, Fig.~\ref{fig:obs_spectra} makes clear the difficulty in distinguishing between these scenarios in a statistically significant way.

\par
This pattern is clearly broken by the best-fit scenario, which is most similar to the $\alpha+\beta = 0$ family of scenarios despite its having $\alpha+\beta \simeq 1.75$. This is due to its significant escaping proton flux and large value of $\alpha$, so that this proton flux is at high energies relative to nuclei. This can be seen explicitly in Figs.~\ref{fig:sibyll_esc_spectra_BF} and~\ref{fig:epos_esc_spectra_BF}. While this feature may not be easily accessible through the high-level information provided by the UHECR spectrum, $\langle X_\mathrm{max} \rangle$, and $\sigma(X_\mathrm{max})$ data, its other signatures will be discussed in the following Section.

\par
We note that the results discussed here may depend significantly on the assumed EBL model, photodisintegration cross-sections, and source evolution. Further study is needed to assess the size of these effects and the generality of our results beyond the scenarios we have studied. However, we leave this to future work as our purpose here is to assess whether there is clear evidence for a Peters cycle under a typical set of reasonable assumptions.

\section{Signatures of alternative scenarios}
\label{sec:4}

\par
Given the difficulty of distinguishing a Peters cycle from alternative scenarios using high-level information like the UHECR spectrum and composition it is worthwhile to explore other signatures which might provide a smoking gun to the dominant process in the universe. In principle, accurate measurements of the spectra of each mass group in the UHECR spectrum would directly access information about $\alpha$ and $\beta$, but this level of mass separation is difficult above the ankle since it would require separation between nuclei of a single mass group. If data below the ankle is also included, such a mass separation is possible~\cite{PierreAuger:2023xfc} but it is not possible to distinguish whether the resulting maximum energy scalings are due to an alternative scenario or a superposition of source populations. However, other signals may provide different means to distinguish between a Peters cycle and alternative scenarios.

\par
First, let us consider the case where there is a substantial proton flux escaping the source. For small values of $\alpha$, this component will peak at low energies relative to nuclei -- a factor of $2^{-\alpha}$ below the energy-per-nucleon of nuclei for the Peters cycle family of scenarios. At such low energies, this proton component likely will be deep into the spectrum's Galactic-to-extragalactic transition. While this lower energy than expected proton component compared to a Peters cycle would be an indication that an alternative scenario is at work, such a signal may be difficult to distinguish from both the protons from photodisintegrated nuclei or a second source population below the ankle.

\par
For large values of $\alpha$, the proton component will peak at higher than expected energies -- a factor of $2^\alpha$ above the energy-per-nucleon of nuclei for the Peters cycle family of scenarios. This would imply that a significant proton component exists in the spectrum at peaking above heavier nuclei, which can result in an increase in $\sigma(X_\mathrm{max})$ at high energies (see Fig.~\ref{fig:obs_spectra}). Additionally, for large enough values of $\alpha$, this proton component will extend beyond the GZK threshold and, therefore, produce a flux of cosmogenic neutrinos at ${\sim}$~EeV energies. Good measurements of the proton fraction throughout the spectrum or observation of cosmogenic neutrinos will probe whether such a component exists and measurement of its peak energy will constrain the value of $\alpha$.

\par
Second, let us consider the case where there is no substantial proton flux escaping the source. In this case, a proton flux will still arrive at Earth due to UHECR photodisintegration interactions with the CMB and extragalactic background light (EBL).\footnote{We assume an approximately continuous distribution of sources with a separation between them which is much smaller than all characteristic propagation lengths, and so the observed spectrum of UHECRs has a universal form, independent of the mode of propagation~\cite{Aloisio:2004jda}. This implies that photodisintegrated protons are still guaranteed to arrive at Earth, despite deflections in the Galactic magnetic field.} These interactions preserve the energy-per-nucleon of the primary CR, so that:
\begin{align}
    E_{A\mathrm{PD}}^p = \frac{E^A}{A}~.
\end{align}
This implies that the peak in the spectrum of photodisintegrated protons from primary CRs of mass $A$ will be given by:
\begin{align}
    E_{A\mathrm{PD, max}}^p = 2^{-\alpha} E_0 A^{\alpha+\beta-1}~.
\end{align}
In the case of the Peters cycle family of scenarios, where $\alpha+\beta=1$, this implies that the spectra of photodisintegrated proton peak at an energy $2^{-\alpha} E_0$ irrespective of their parent CR's mass $A$. Moreover, the peak energy of any primary proton spectrum will be the same as the photodisintegrated spectrum's up to a factor of $2^{-\alpha}$. Distinguishing a Peters cycle from alternative scenarios in its family is very difficult and can only be probed by through this factor of $2^{-\alpha}$ difference in the peak energies of the primary proton component and photodisintegrated proton component.

\par
However, this is not true for alternative families of scenarios where $\alpha+\beta\neq 1$, and the ratio between the peaks in the spectra of photodisintegrated protons from CRs of mass $A$ and mass $A'$ will be given by:
\begin{equation}
    \frac{E^p_{A\mathrm{PD,max}}}{E^p_{A'\mathrm{PD,max}}} = \left( \frac{A}{A'} \right)^{\alpha+\beta-1}~.
\end{equation}
Therefore, a signature of alternative scenarios outside the Peters cycle family is a multi-peaked proton component in the UHECR spectrum. While it may be difficult to resolve these separate peaks, this signature also implies an extended proton spectrum throughout the entire UHECR spectrum, which is unexpected for scenarios in the Peters cycle family. This extended proton component may be more easily detected than resolve the peak energy of spectrum of each mass group, and can result in larger-than-expected values of $\sigma(X_\mathrm{max})$ at high energies.

\par
An important caveat, however, is that an alternative scenario to a Peters cycle may be difficult to distinguish from a second UHECR source population which produces a substantial high-energy proton flux (this possibility has been explored in a number of studies, including~\cite{vanVliet:2019nse,Muzio:2019leu,Muzio:2023skc,Ehlert:2023btz}). Such a source population could mimic alternative scenarios by producing cosmogenic neutrinos or a larger than expected proton fraction through the spectrum. If the fluxes of heavier nuclei produced by this source population are not large enough to be detected, it may be impossible to distinguish between the two.

\section{Conclusions}
\label{sec:5}

\par
In this study we explored the degree to which observations of the ultrahigh energy cosmic ray spectrum and composition above the ankle favor a Peters cycle over alternative scenarios for spectra escaping sources. Such alternatives are motivated by energy loss and beyond the Standard Model processes which imprint particular scalings between the maximum energy of nuclei and their mass and charge. We find that alternative scenarios explain the UHECR data above the ankle better than a Peters cycle, regardless of the hadronic interaction model or systematic data shifts assumed, for the scenarios we explored. This result raises the possibility that a Peters cycle is not realized in any energy range of the observed UHECR spectrum.

\par
We investigated the observational signatures which might be used to further discriminate alternative scenarios from a Peters cycle. These include unexpected scalings of the peak energy of different mass groups at Earth, a substantial GZK neutrino flux, an extended proton component across the spectrum's full energy range, or an unexpected proton component at the highest energies (which could result in a large $\sigma(X_\mathrm{max})$ and therefore may be constrained by deep neural network-based $X_\mathrm{max}$ measurements~\cite{PierreAuger:2023gmj}). However, some of these signatures are difficult to distinguish from an additional UHECR source population below the ankle or one producing a substantial high-energy proton flux above it.

\par
We emphasize that alternative scenarios to a Peters cycle represent an exciting observational opportunity. In particular, further constraints on these scenarios will not only significantly reduce the theoretical uncertainties in UHECR modeling and open a window to the conditions inside UHECR sources, but they will also enable UHECR data to directly constrain new physics processes.

\section*{Acknowledgments}

The work of L.A.A. is supported by the U.S. National Science
Foundation (NSF Grant PHY-2112527). The research of M.S.M. is supported by the NSF MPS-Ascend Postdoctoral Award \#2138121.

\bibliography{mua}

\appendix

\section{Results for systematic shifts of data} \label{app:sysShifts}

\par
\begin{figure*}[htbp!]
    \centering
    \begin{minipage}{0.32\linewidth}
	  \centering
      \subfloat[]{\includegraphics[width=\textwidth]{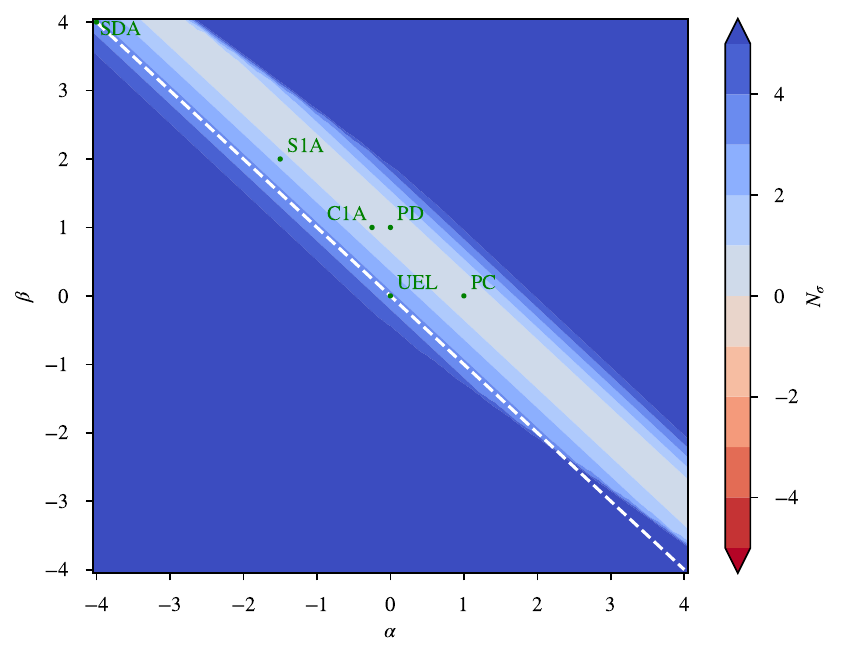}}
    \end{minipage}
    \begin{minipage}{0.32\linewidth}
	  \centering
      \subfloat[]{\includegraphics[width=\textwidth]{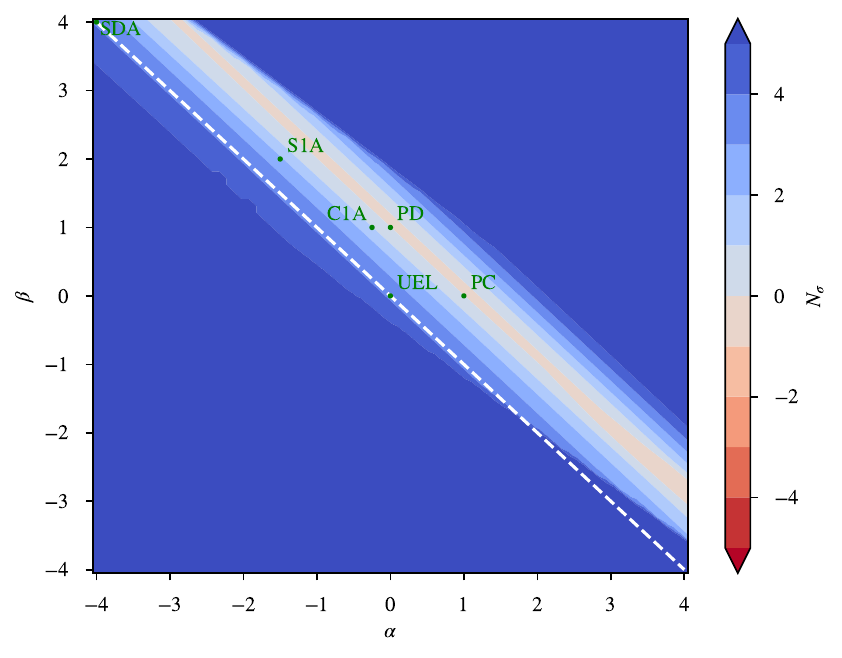}}
    \end{minipage}
    \begin{minipage}{0.32\linewidth}
	  \centering
      \subfloat[]{\includegraphics[width=\textwidth]{AZRmaxDependence_noSrc_sibyll_hiE_Eshift1_Xshift-1_NsigmaRel.pdf}}
    \end{minipage}

    \begin{minipage}{0.32\linewidth}
	  \centering
      \subfloat[]{\includegraphics[width=\textwidth]{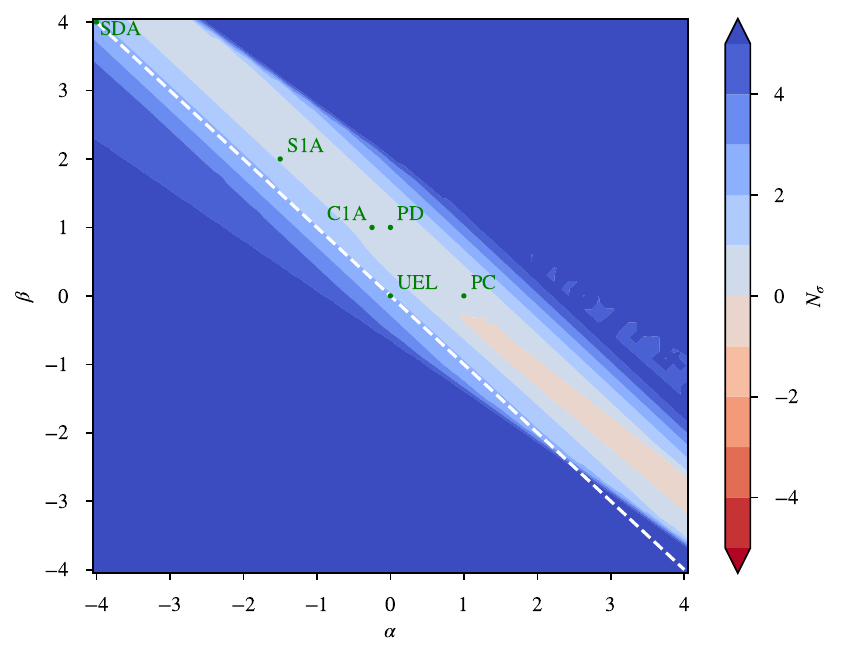}}
    \end{minipage}
    \begin{minipage}{0.32\linewidth}
	  \centering
      \subfloat[]{\includegraphics[width=\textwidth]{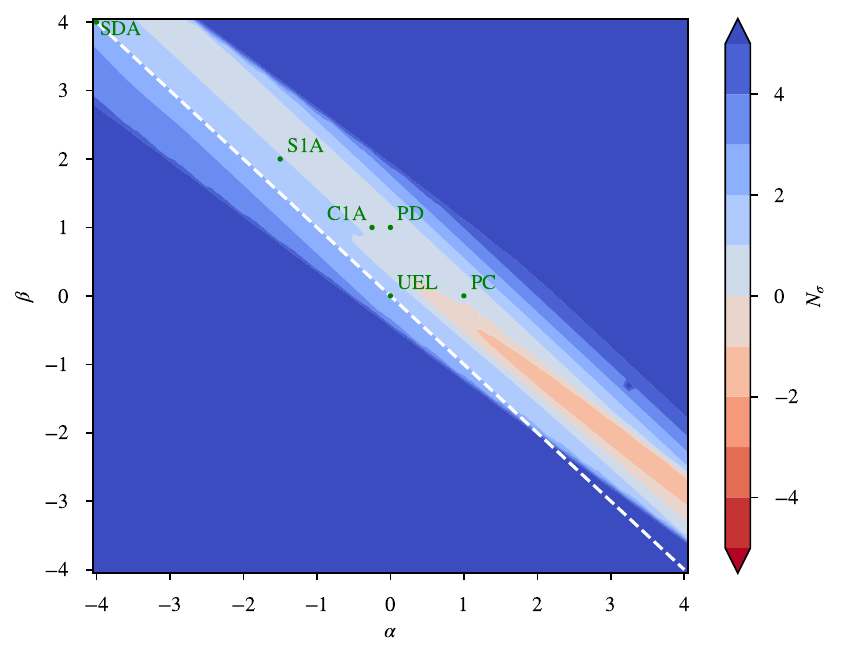}}
    \end{minipage}
    \begin{minipage}{0.32\linewidth}
	  \centering
      \subfloat[]{\includegraphics[width=\textwidth]{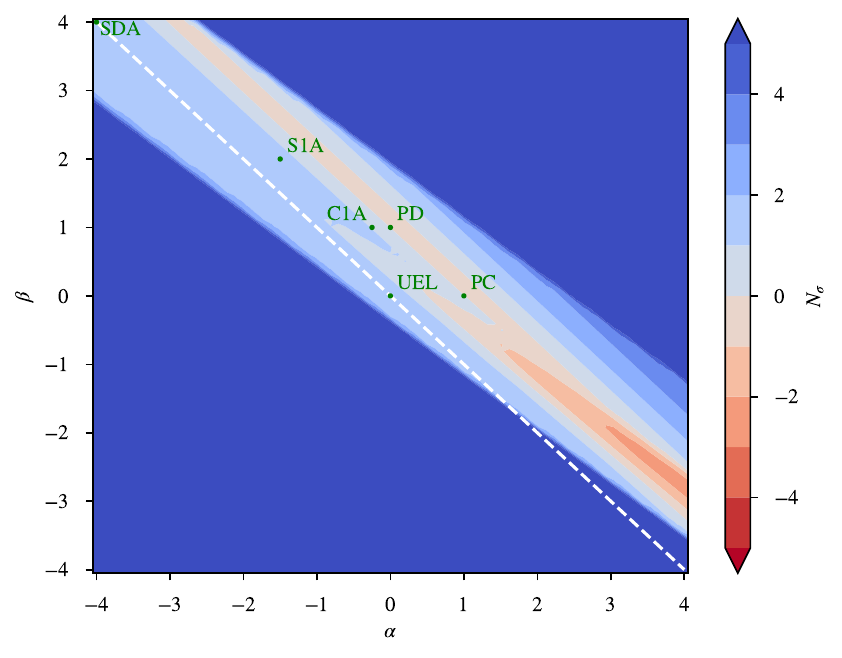}}
    \end{minipage}

    \begin{minipage}{0.32\linewidth}
	  \centering
      \subfloat[]{\includegraphics[width=\textwidth]{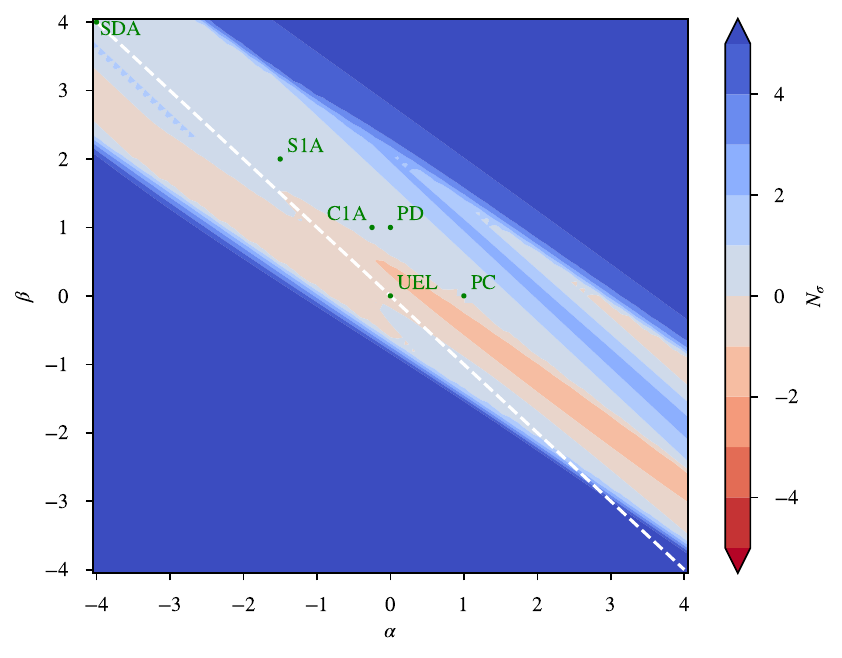}}
    \end{minipage}
    \begin{minipage}{0.32\linewidth}
	  \centering
      \subfloat[]{\includegraphics[width=\textwidth]{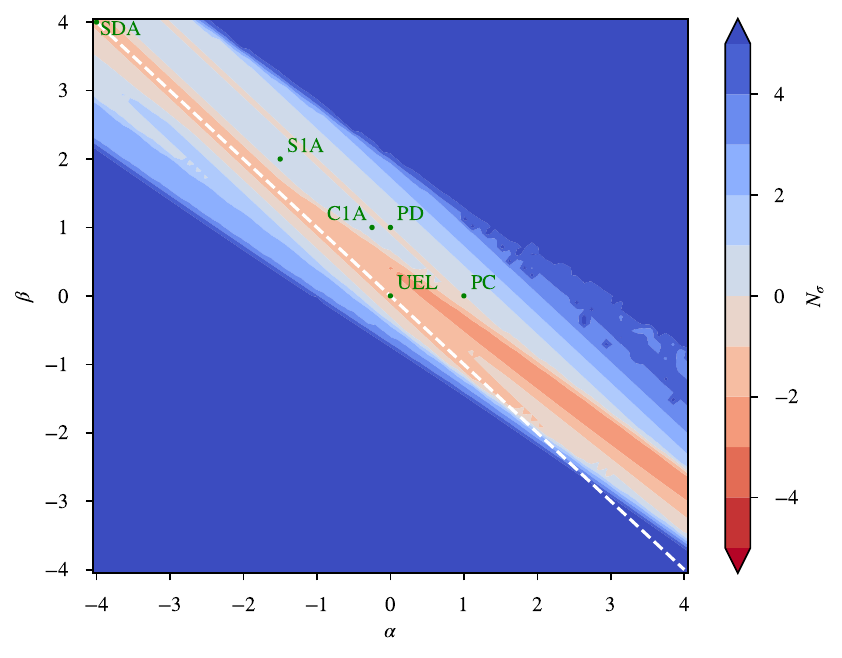}}
    \end{minipage}
    \begin{minipage}{0.32\linewidth}
	  \centering
      \subfloat[]{\includegraphics[width=\textwidth]{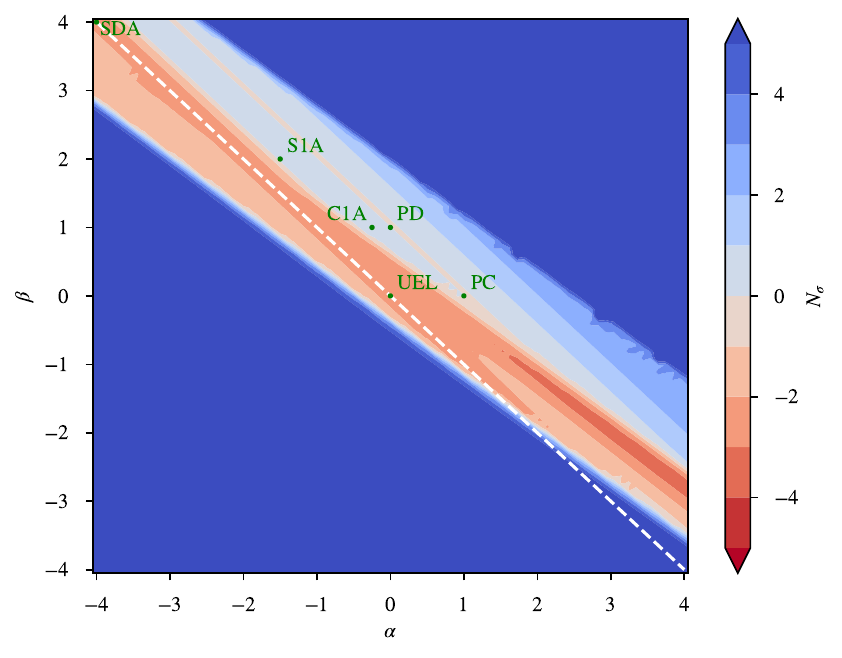}}
    \end{minipage}
	\caption{Change in quality of fit to the UHECR spectrum and composition relative to a Peters cycle when considering different systematic shifts of the data. We consider \textsc{Sibyll2.3d} shifting the data by $\mathrm{dlg}E=-0.1/0/+0.1$ (left/middle/right columns) and $-1/0/+1\sigma_X$ (top/middle/bottom rows).The Peters cycle (PC) and a number of alternative scenarios (green dots) are highlighted: a photodisintegration-limited spectrum (PD), a synchrotron-limited diffusion accelerated spectrum (SDA), a synchrotron-limited one-shot accelerated spectrum (S1A), a curvature radiation-limited one-shot accelerated spectrum (C1A), and a universal energy loss spectrum (UEL).}
	\label{fig:sibyll_shifts}
\end{figure*}
\begin{figure*}[htbp!]
    \centering
    \begin{minipage}{0.32\linewidth}
	  \centering
      \subfloat[]{\includegraphics[width=\textwidth]{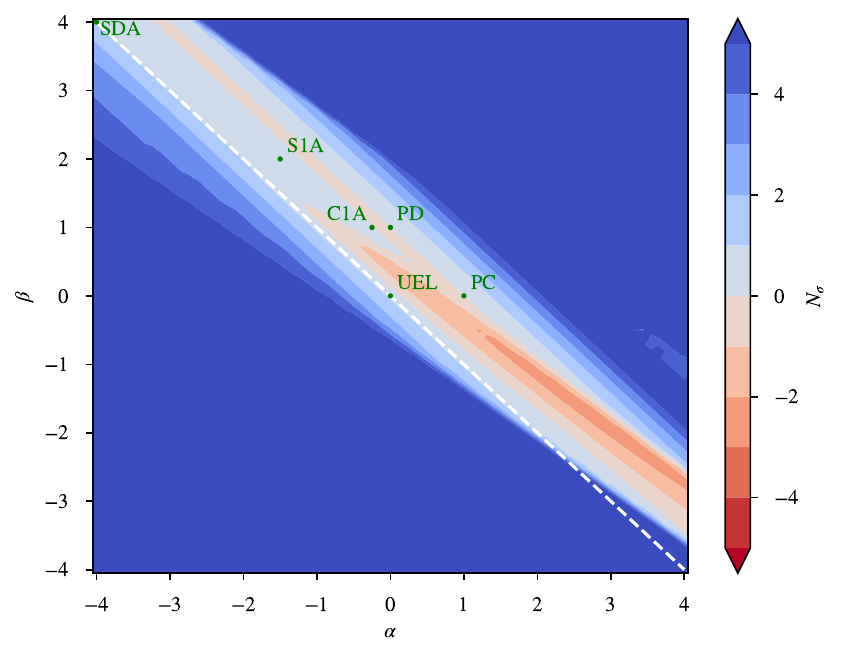}}
    \end{minipage}
    \begin{minipage}{0.32\linewidth}
	  \centering
      \subfloat[]{\includegraphics[width=\textwidth]{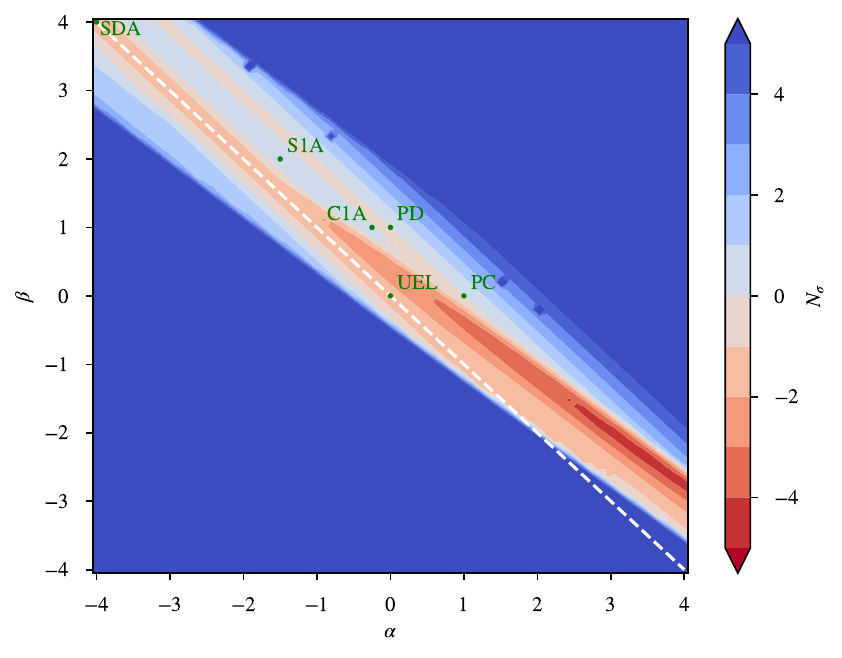}}
    \end{minipage}
    \begin{minipage}{0.32\linewidth}
	  \centering
      \subfloat[]{\includegraphics[width=\textwidth]{AZRmaxDependence_noSrc_epos_hiE_Eshift1_Xshift-1_NsigmaRel.pdf}}
    \end{minipage}

    \begin{minipage}{0.32\linewidth}
	  \centering
      \subfloat[]{\includegraphics[width=\textwidth]{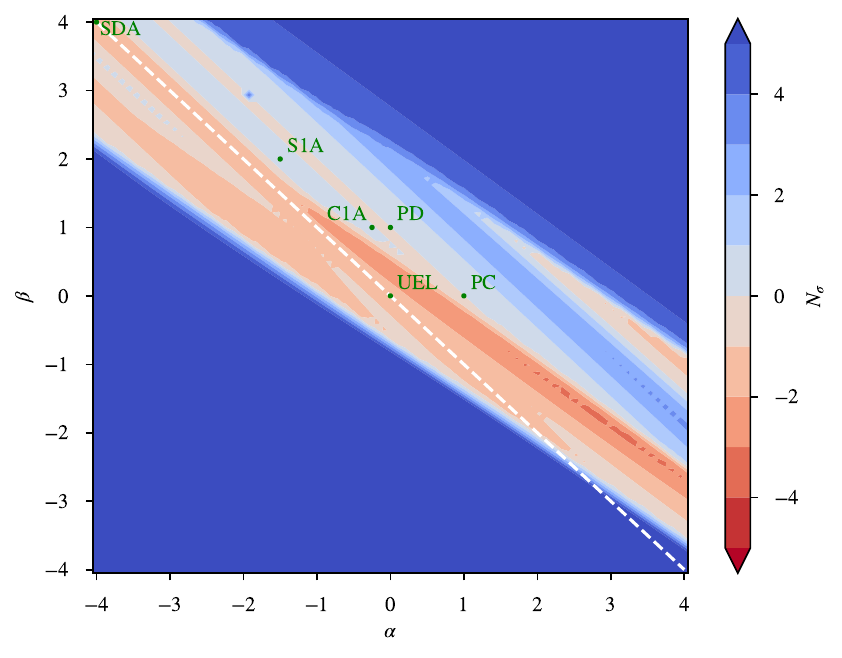}}
    \end{minipage}
    \begin{minipage}{0.32\linewidth}
	  \centering
      \subfloat[]{\includegraphics[width=\textwidth]{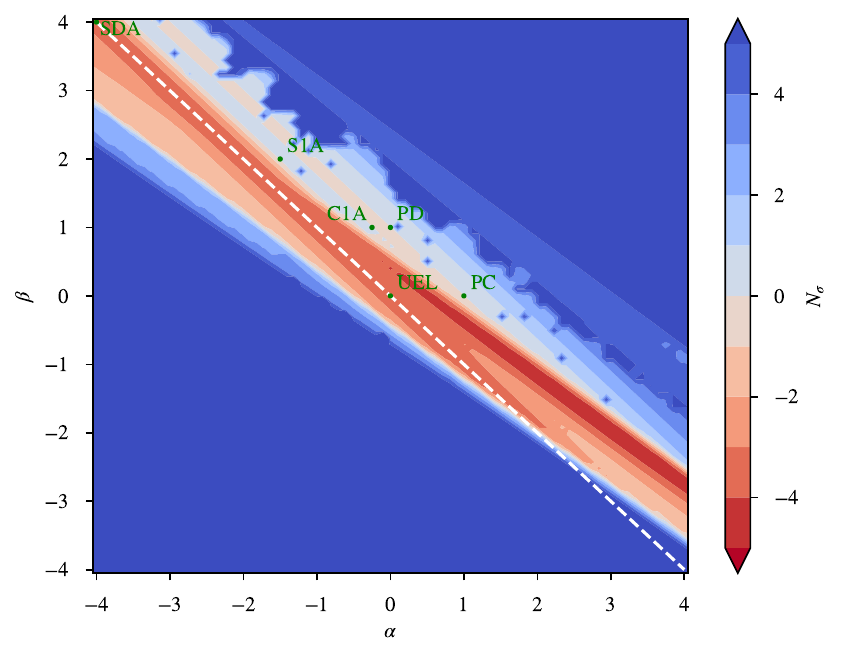}}
    \end{minipage}
    \begin{minipage}{0.32\linewidth}
	  \centering
      \subfloat[]{\includegraphics[width=\textwidth]{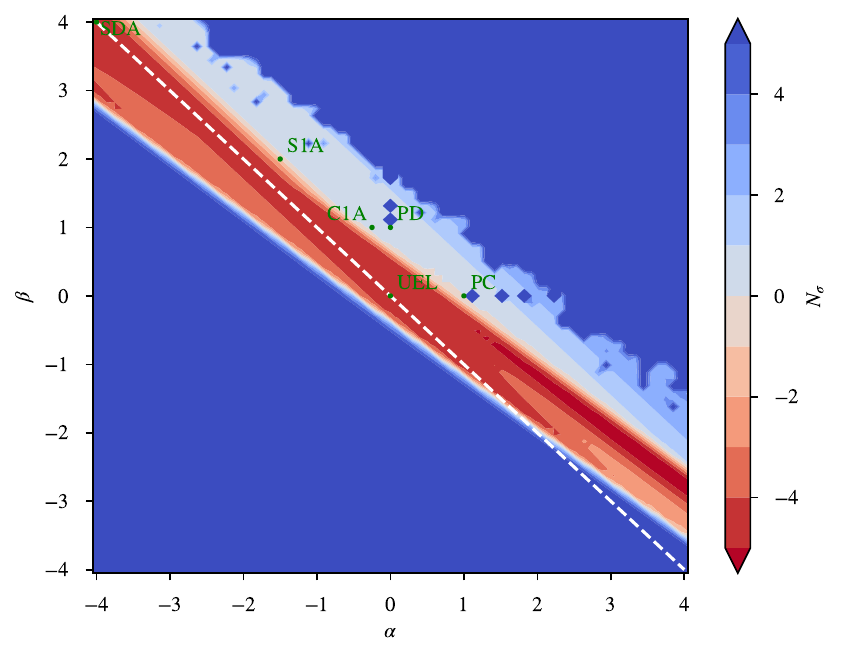}}
    \end{minipage}

    \begin{minipage}{0.32\linewidth}
	  \centering
      \subfloat[]{\includegraphics[width=\textwidth]{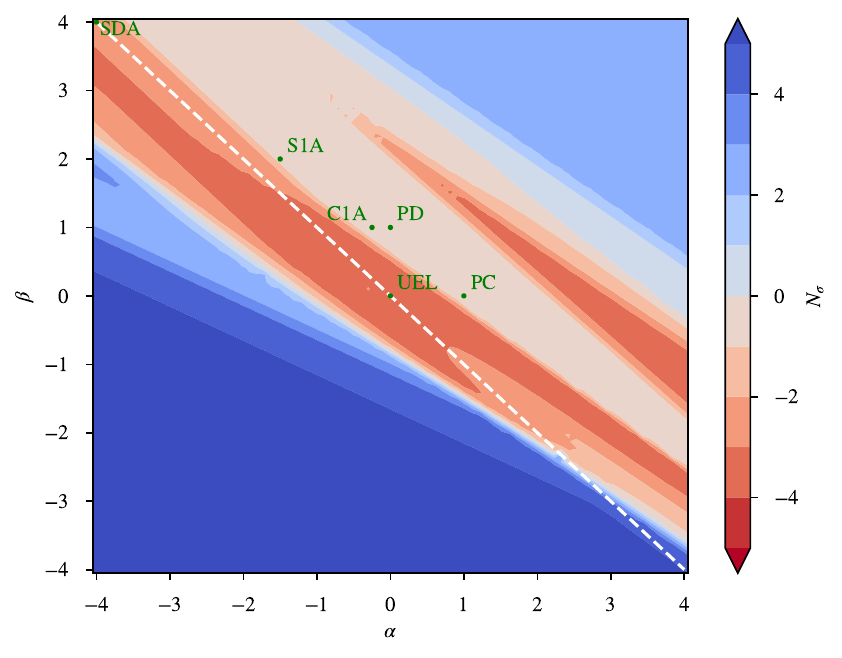}}
    \end{minipage}
    \begin{minipage}{0.32\linewidth}
        \centering
        \subfloat[]{\includegraphics[width=\textwidth]{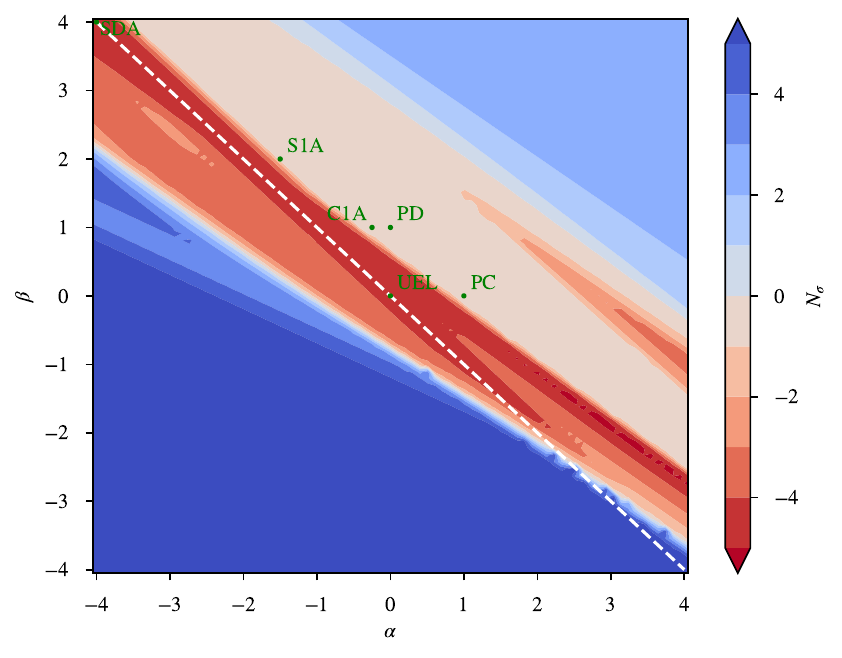}}
    \end{minipage}
    \begin{minipage}{0.32\linewidth}
	  \centering
      \subfloat[]{\includegraphics[width=\textwidth]{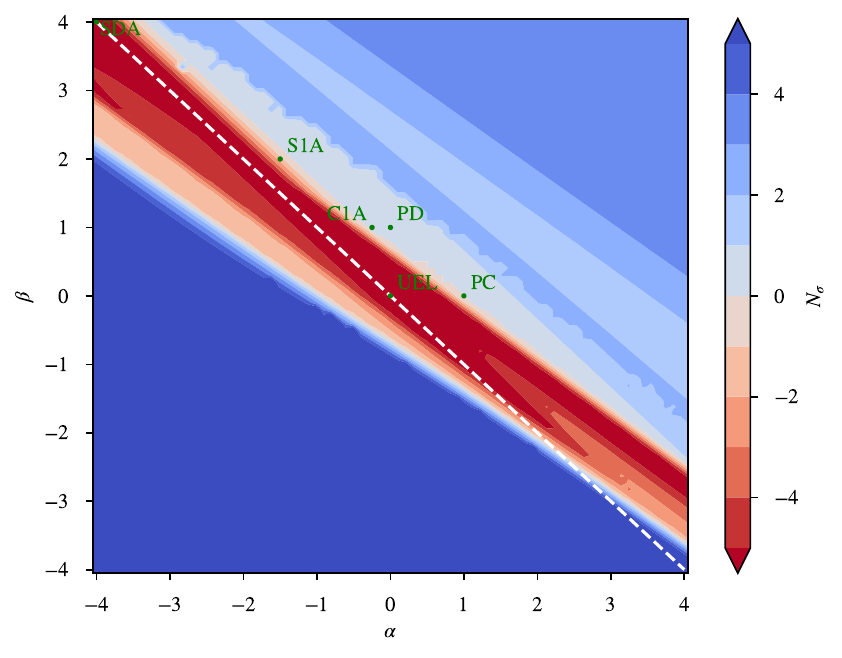}}
    \end{minipage}
	\caption{Same as Fig.~\ref{fig:sibyll_shifts} for \textsc{Epos-LHC}.}
	\label{fig:epos_shifts}
\end{figure*}

Figures~\ref{fig:sibyll_shifts} and~\ref{fig:epos_shifts} show the effect of systematically shifting the spectrum and composition data under both hadronic interaction models considered. We consider all combinations shifting the energy scale by $\mathrm{dlg}E=-0.1/0/+0.1$ and shifting the $\langle X_\mathrm{max} \rangle$ by $-1/0/+1\sigma_X$. As can be seen, these shifts have a non-trivial impact on the particular regions of parameter space which best describe the data. However, generically it is the case that no matter the systematic shifts on consider alternative scenarios describe the data as well or better than a Peters cycle.

\par
If one has a strong theoretical prior towards a Peters cycle or a particular alternative scenario, this will strongly effect the systematic data shifts one favors. Importantly, we note that the overall quality of fit (i.e. the value of the $\chi^2$) depends strongly on the systematic shifts chose regardless of the $(\alpha, \beta)$ value one considers.

\section{Alternative scenario escaping spectra}\label{app:esc_spectra}

\par
\begin{figure*}
    \centering
    \begin{minipage}{0.32\linewidth}
	  \centering
        \subfloat[]{\includegraphics[width=\textwidth]{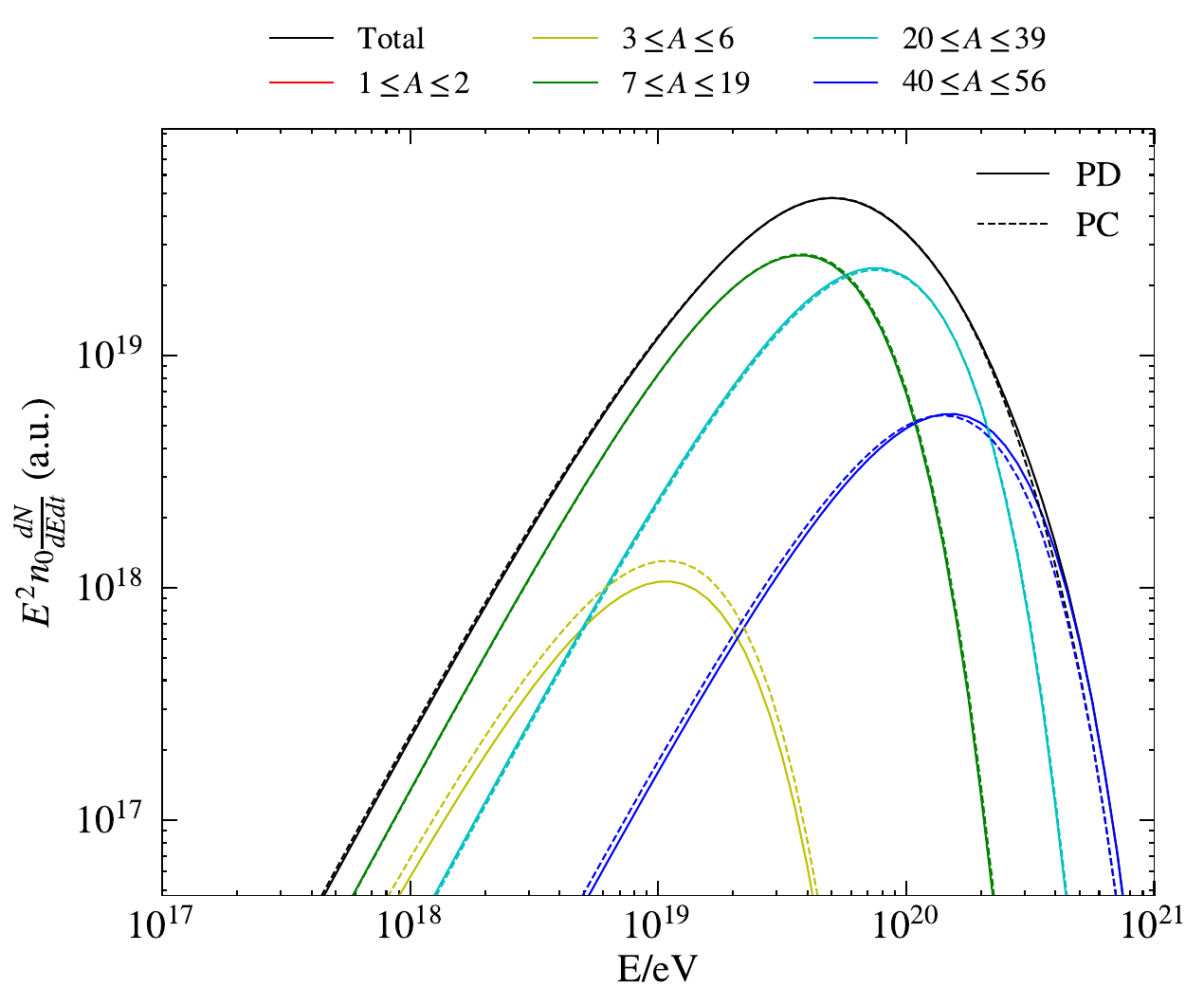}}
    \end{minipage}
    \begin{minipage}{0.32\linewidth}
	  \centering
        \subfloat[]{\includegraphics[width=\textwidth]{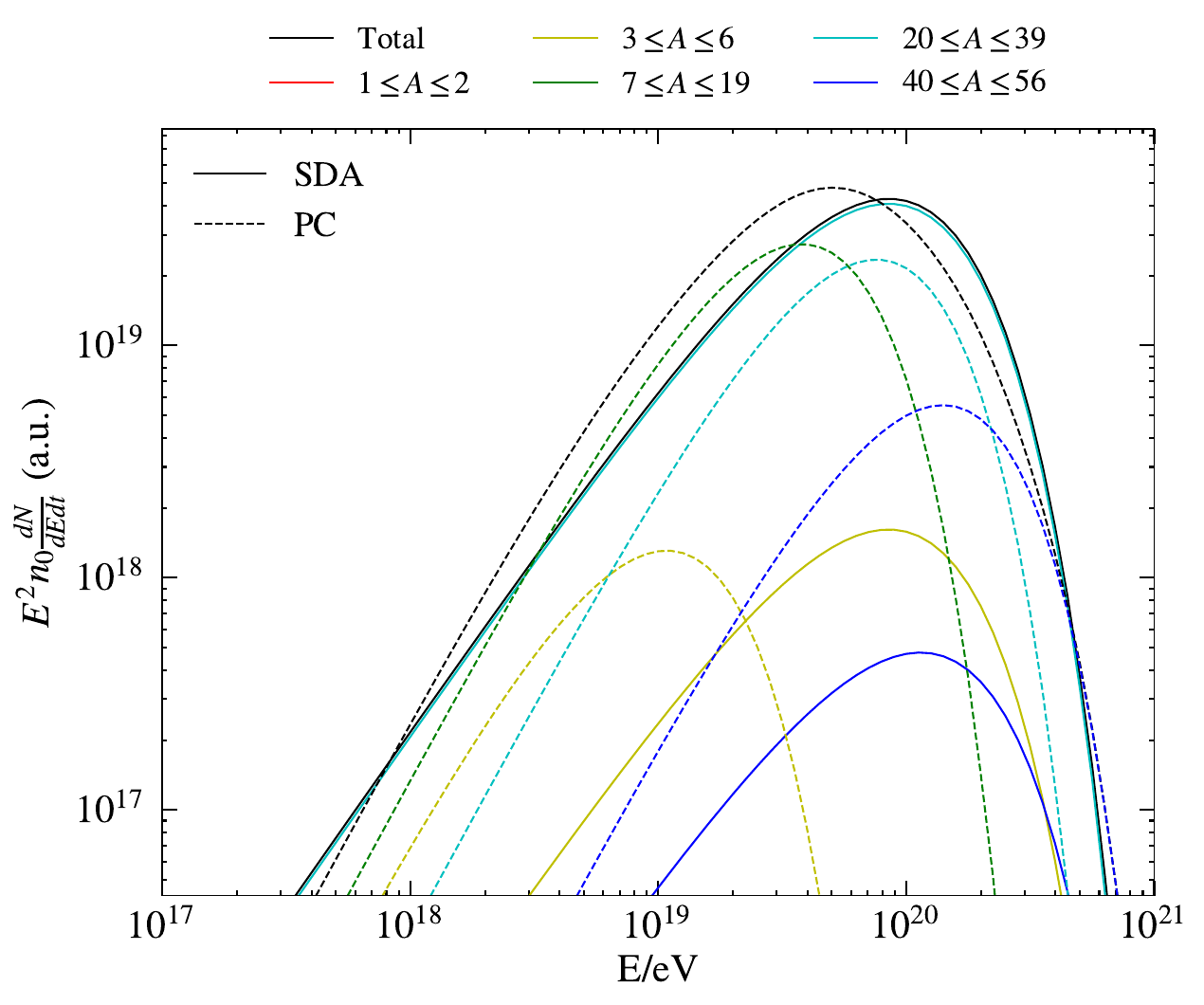}}
    \end{minipage}
    \begin{minipage}{0.32\linewidth}
	  \centering
        \subfloat[]{\includegraphics[width=\textwidth]{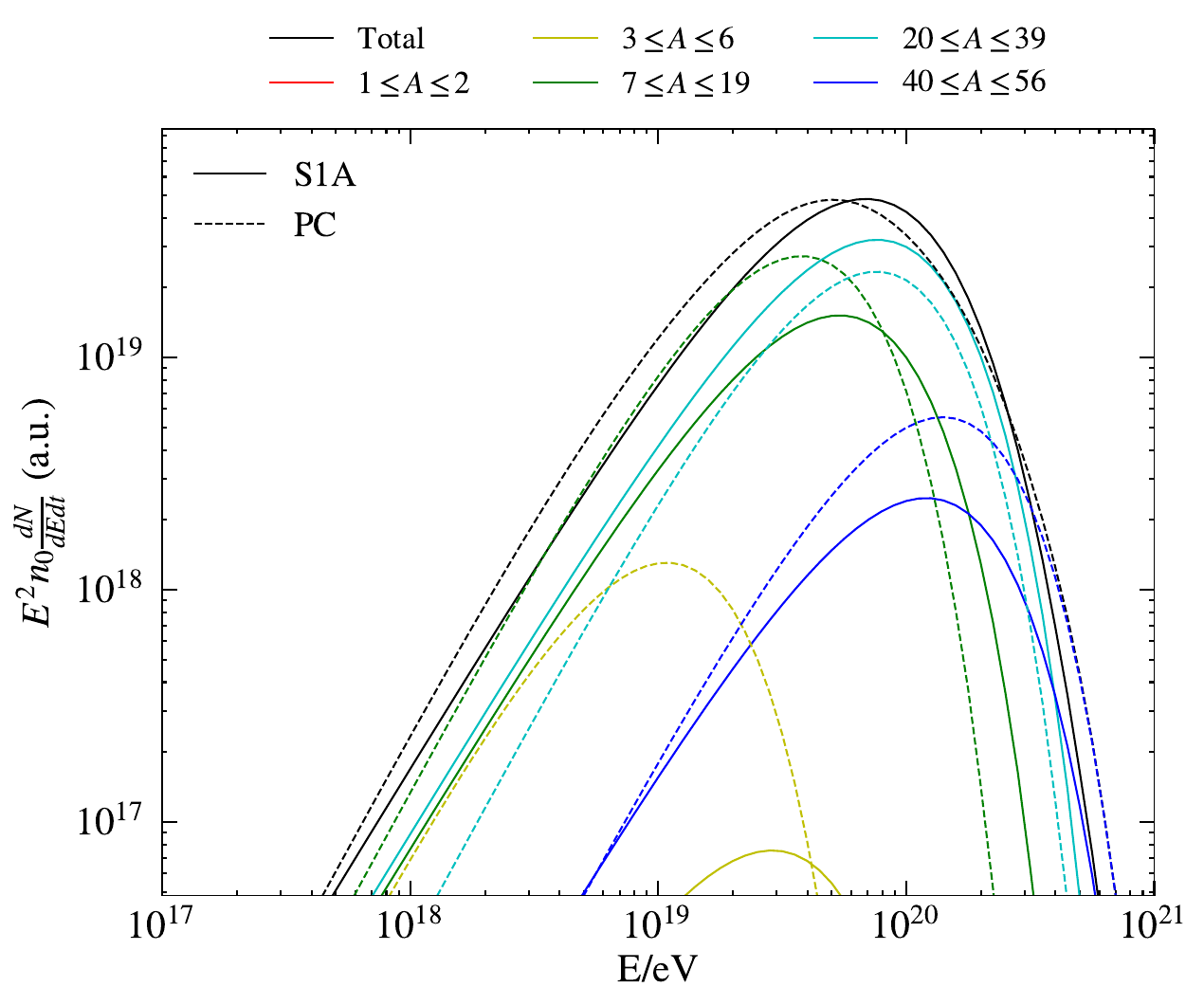}}
    \end{minipage}
    \begin{minipage}{0.32\linewidth}
	  \centering
        \subfloat[]{\includegraphics[width=\textwidth]{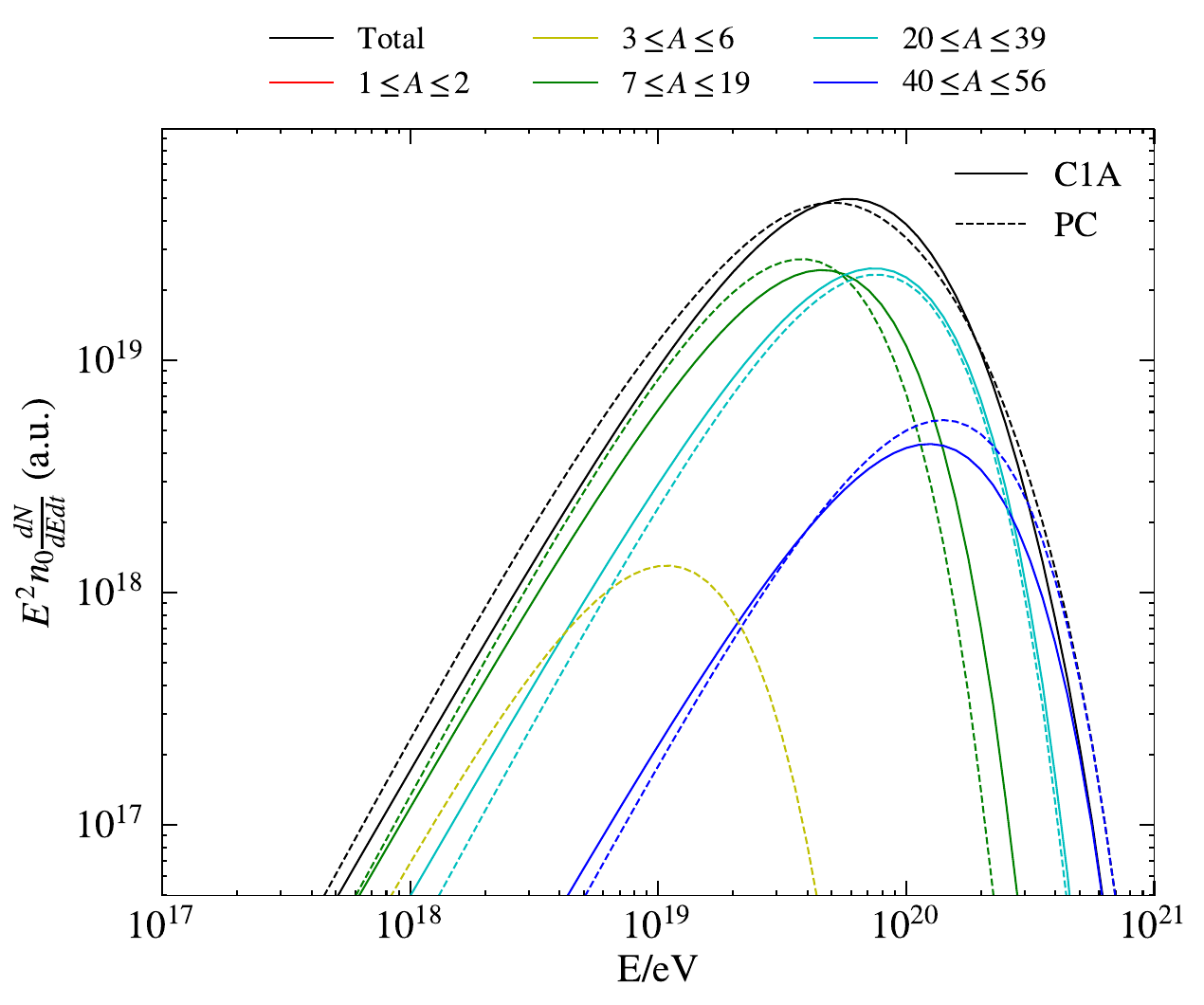}}
    \end{minipage}
    \begin{minipage}{0.32\linewidth}
	  \centering
        \subfloat[]{\includegraphics[width=\textwidth]{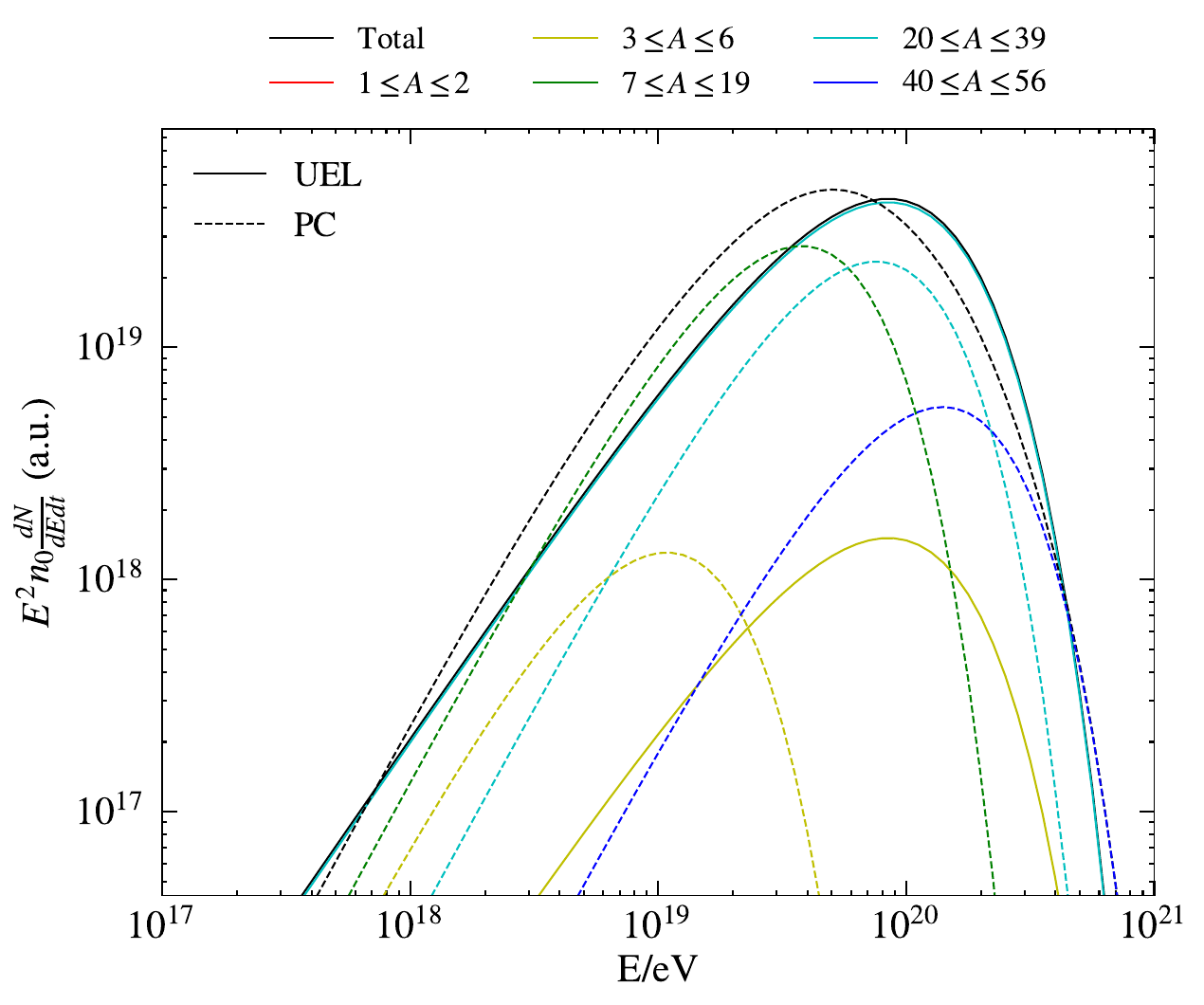}}
    \end{minipage}
    \begin{minipage}{0.32\linewidth}
	  \centering
        \subfloat[\label{fig:sibyll_esc_spectra_BF}]{\includegraphics[width=\textwidth]{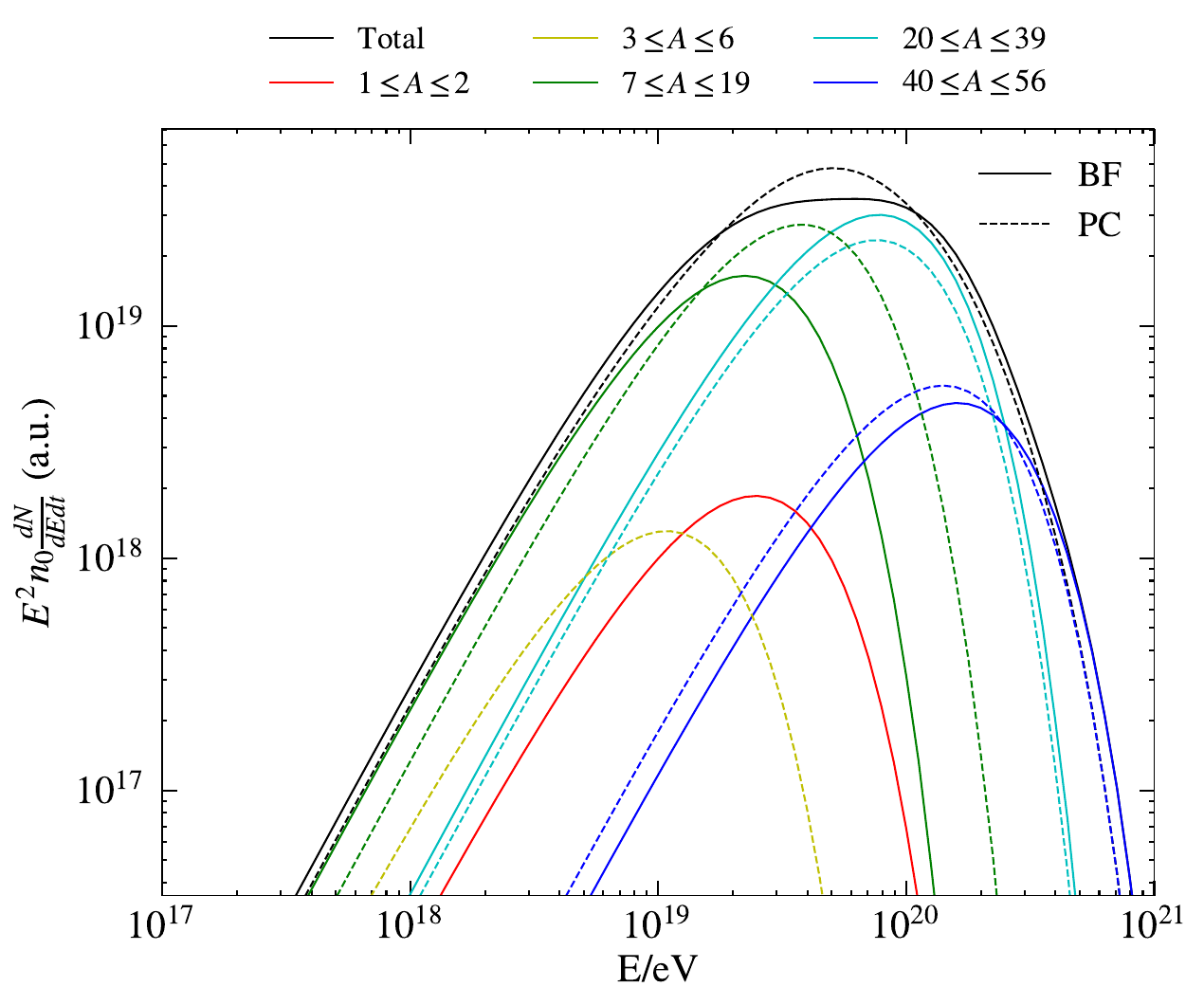}}
    \end{minipage}

    \caption{Best-fit escaping spectra for non-Peters cycle scenarios (solid lines) compared to those for a Peters cycle (dashed lines). Spectra are broken down by mass group (colored lines). All plots are for \textsc{Sibyll2.3d} with data shifted by $\mathrm{dlg}E=+0.1$ and $-1\sigma_X$.}
    \label{fig:sibyll_esc_spectra}
\end{figure*}
\begin{figure*}
    \centering
    \begin{minipage}{0.32\linewidth}
	  \centering
        \subfloat[]{\includegraphics[width=\textwidth]{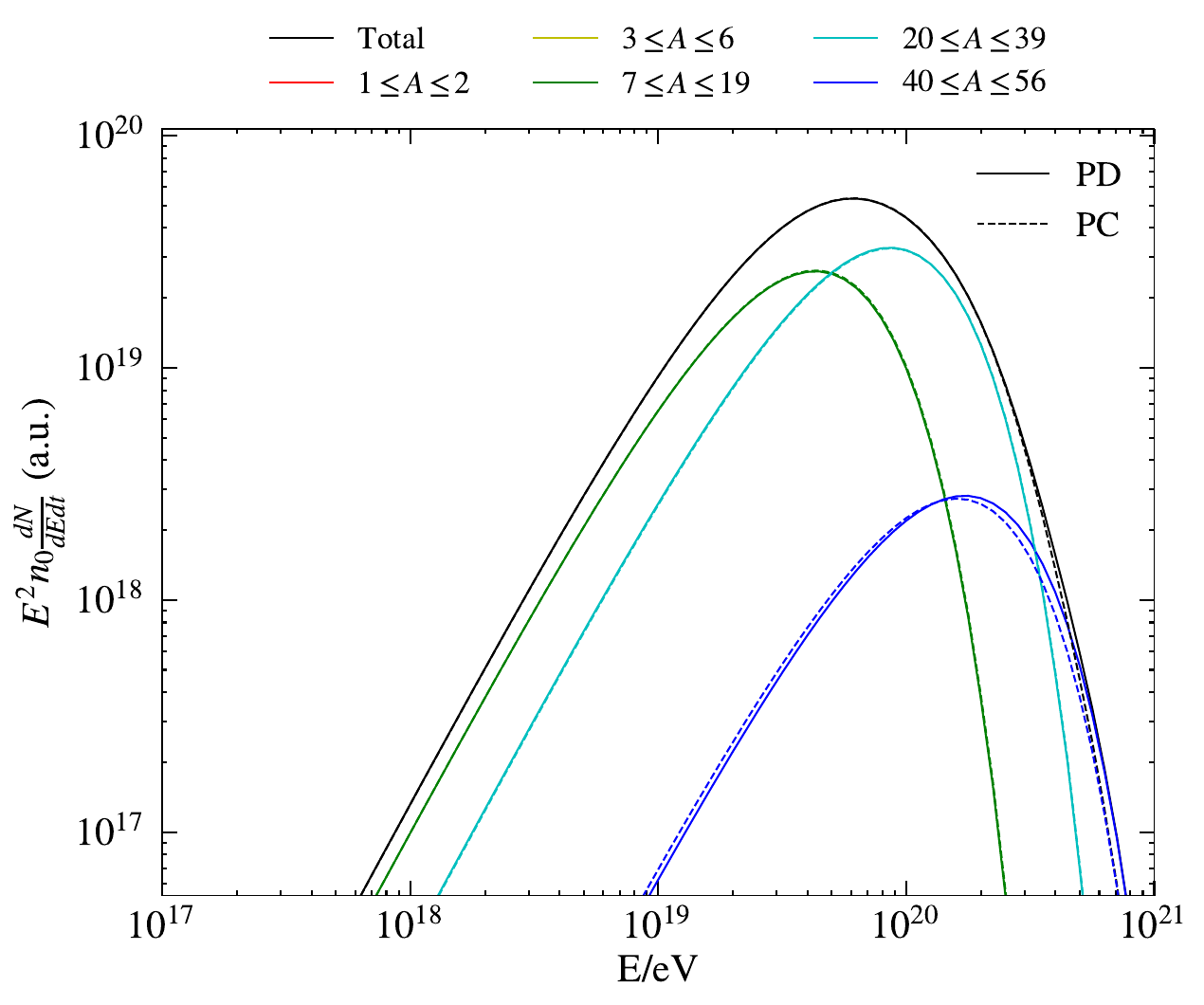}}
    \end{minipage}
    \begin{minipage}{0.32\linewidth}
	  \centering
        \subfloat[]{\includegraphics[width=\textwidth]{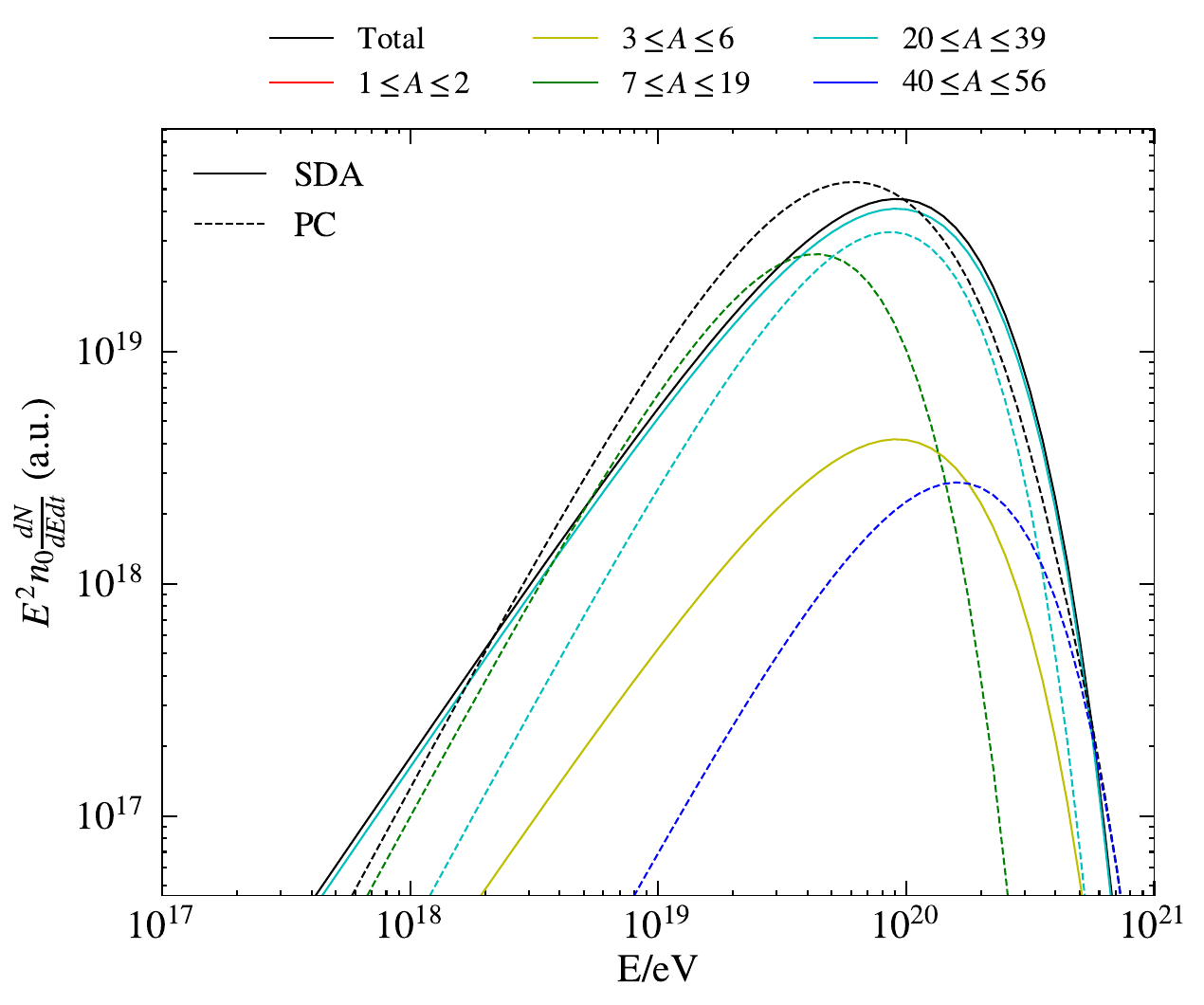}}
    \end{minipage}
    \begin{minipage}{0.32\linewidth}
	  \centering
        \subfloat[]{\includegraphics[width=\textwidth]{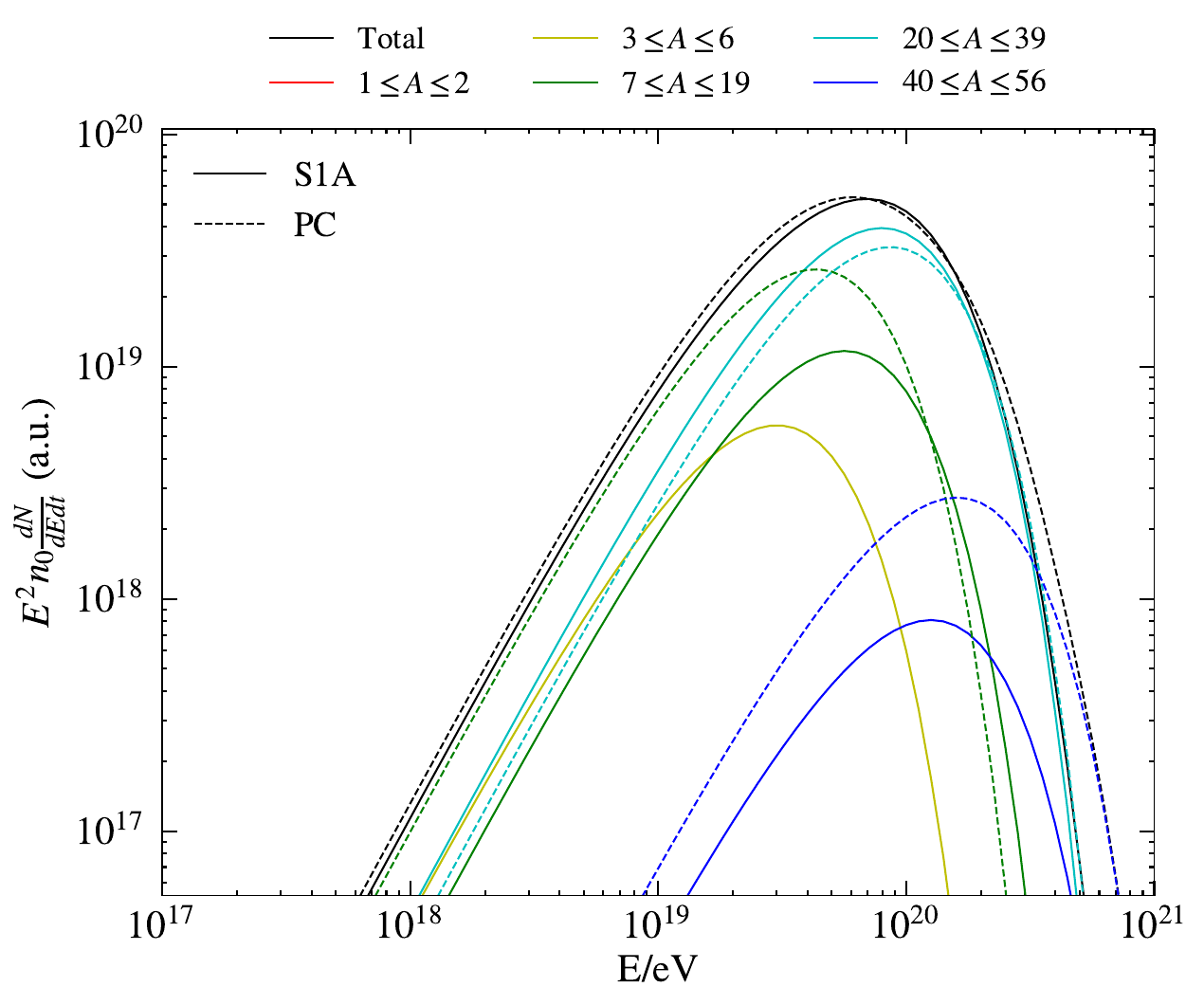}}
    \end{minipage}
    \begin{minipage}{0.32\linewidth}
	  \centering
        \subfloat[]{\includegraphics[width=\textwidth]{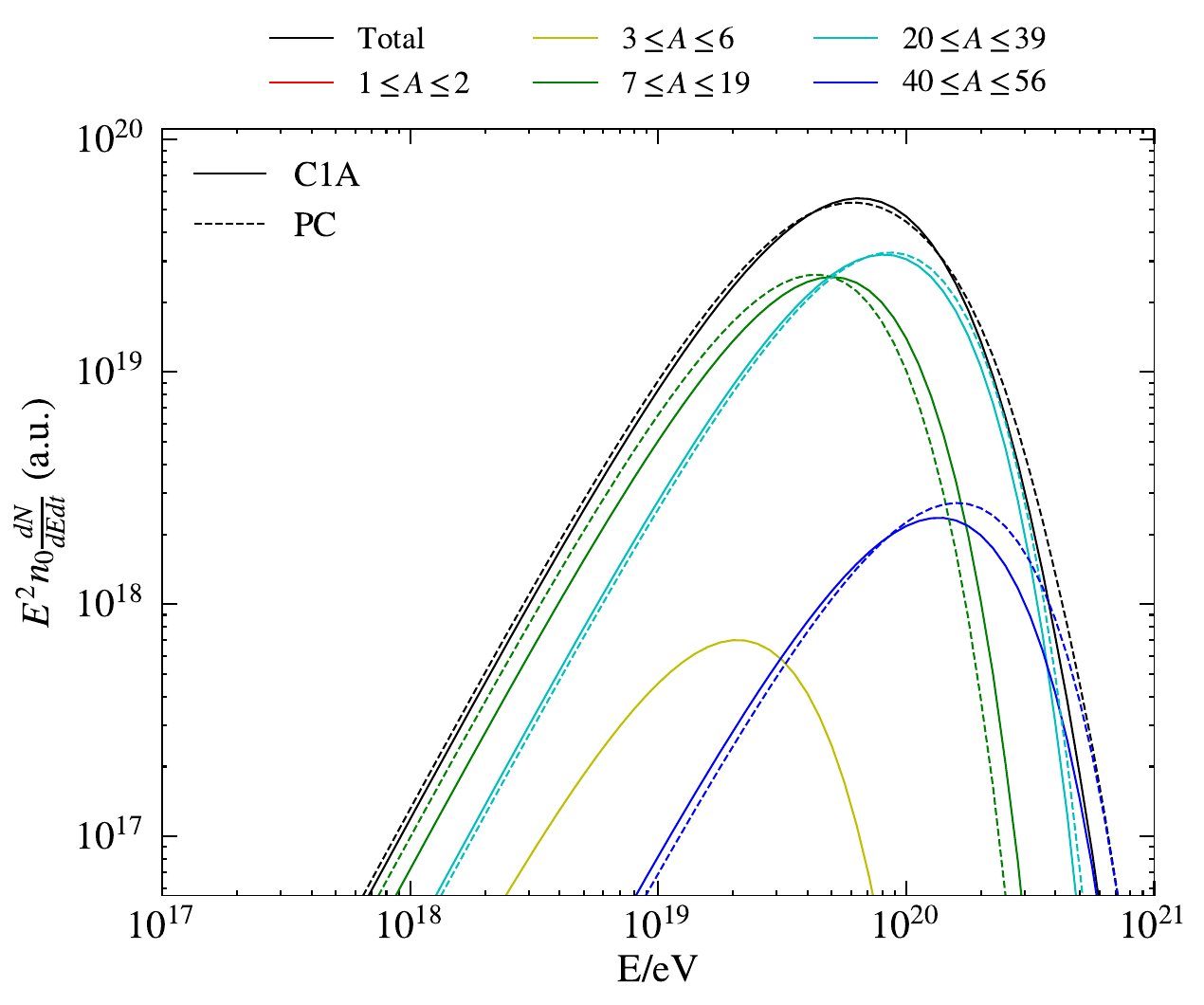}}
    \end{minipage}
    \begin{minipage}{0.32\linewidth}
	  \centering
        \subfloat[]{\includegraphics[width=\textwidth]{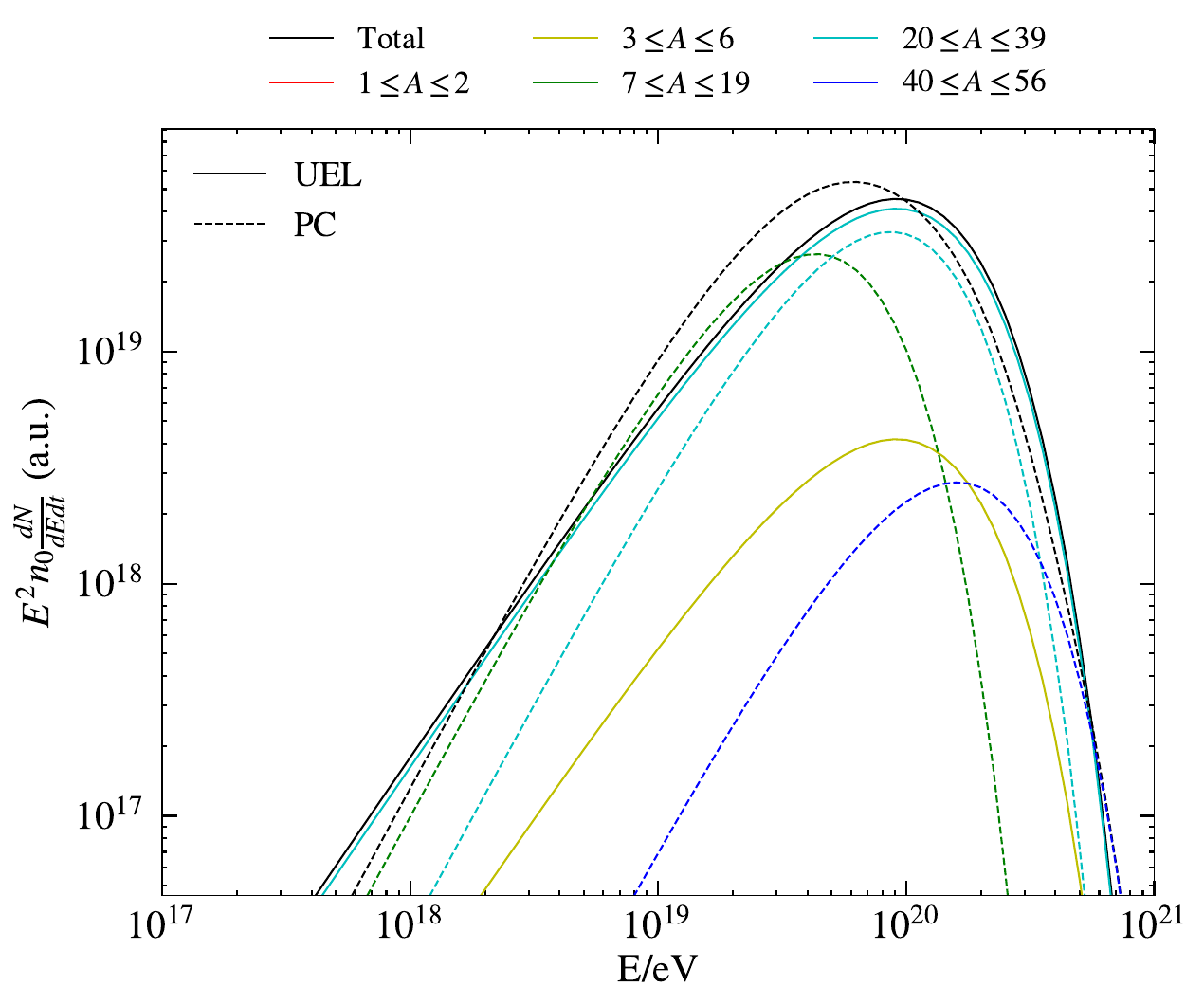}}
    \end{minipage}
    \begin{minipage}{0.32\linewidth}
	  \centering
        \subfloat[\label{fig:epos_esc_spectra_BF}]{\includegraphics[width=\textwidth]{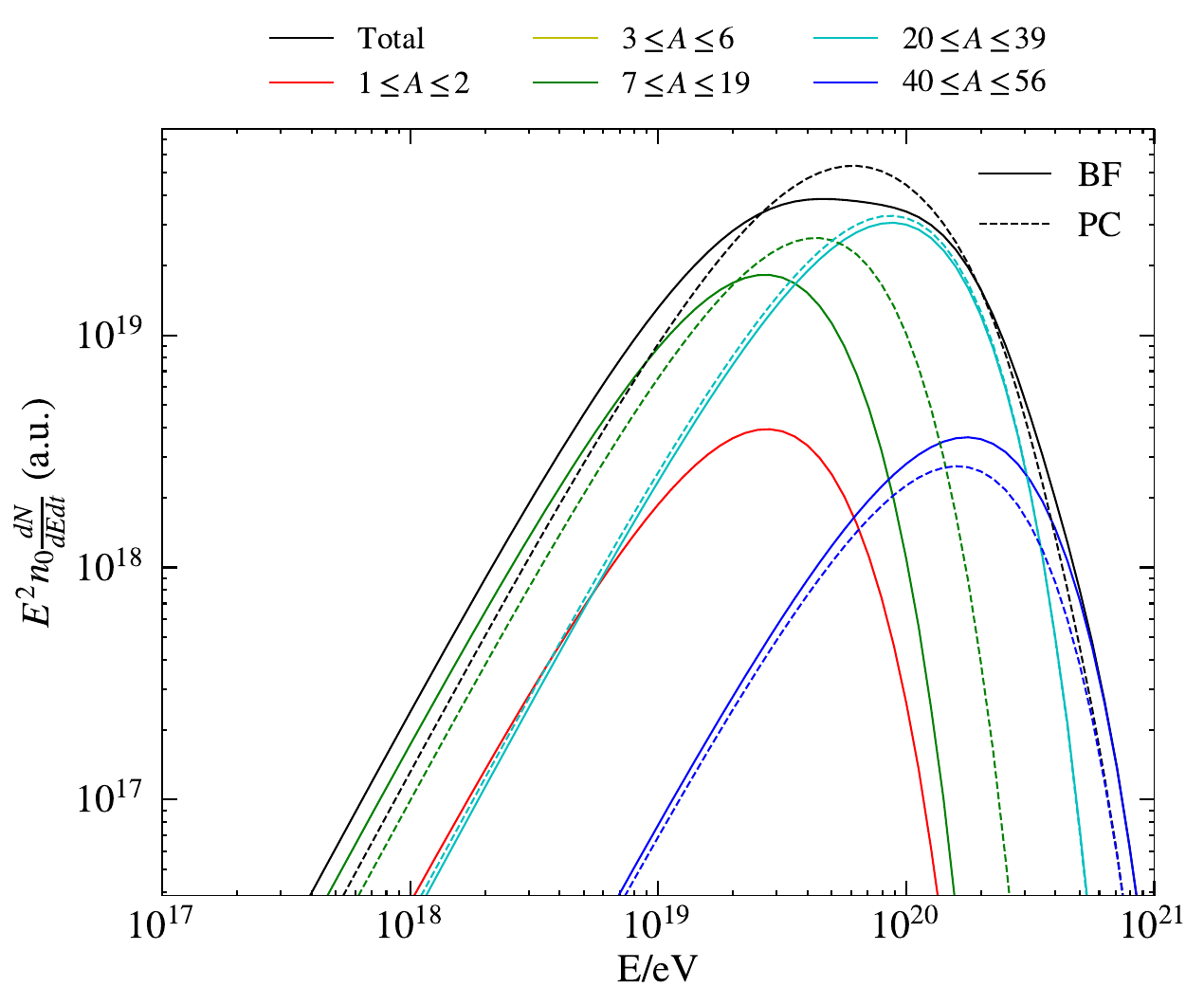}}
    \end{minipage}

    \caption{Same as Fig.~\ref{fig:sibyll_esc_spectra} but for \textsc{Epos-LHC}.}
    \label{fig:epos_esc_spectra}
\end{figure*}

Figures~\ref{fig:sibyll_esc_spectra} and~\ref{fig:epos_esc_spectra} show the best-fit escaping spectra for alternative scenarios compared to a Peters cycle for our benchmark systematic data shifts. These plots underscore the similarity of scenarios belonging to the same $\alpha+\beta$ family: Peters cycle \& photodisintegration ($\alpha+\beta=1$), and synchrotron-limited diffusion acceleration \& universal energy losses ($\alpha+\beta=0$).

\par
The best-fit scenario ($\alpha \simeq 6.75$, $\beta \simeq -5$) also demonstrates how scenarios with a significant proton component and large $\alpha$ result in a proton component which does not obey the normal mass ordering expected in a Peters cycle. In this case, the proton component peaks at similar energies as the CNO component, resulting in a peak in the $\sigma(X_\mathrm{max})$ at high energies (see Fig.~\ref{fig:obs_spectra}).

\end{document}